\documentclass[prb,amsmath,amssymb,reprint]{revtex4-1}

\usepackage{blindtext}
\usepackage{amsmath}
\usepackage{siunitx}
\usepackage{graphicx}
\usepackage{enumerate}

\newcommand{\mb}[1]{\mathbf{#1}}

\usepackage{hyperref}
\hypersetup{
	colorlinks=true,
	linkcolor=blue,    
	urlcolor=blue,
}

\usepackage[capitalise]{cleveref}

\begin{document}
\title{Global optimization of atomistic structure enhanced by machine learning}
	
\author{Malthe K. Bisbo}
\author{Bj{\o}rk Hammer}

\affiliation{
Department of Physics and Astronomy, Aarhus university, DK-8000 Aarhus C, Denmark		
}

\date{\today}

\begin{abstract}
Global Optimization with First-principles Energy Expressions
(GOFEE) is an efficient method for identifying low energy structures
in computationally expensive energy landscapes such as the ones
described by density functional theory (DFT), van der Waals-enabled
DFT, or even methods beyond DFT. GOFEE relies on a machine learned
surrogate model of energies and forces, trained on-the-fly, to explore
configuration space, eliminating the need for expensive relaxations of
all candidate structures using first-principles methods. In this paper
we elaborate on the importance of the use of a Gaussian kernel with
two length scales in the Gaussian Process Regression (GPR) surrogate
model. We further explore the role of the use in GOFEE of the lower
confidence bound for relaxation and selection of candidate
structures. In addition, we present two improvements to the method: 1)
the population generation now relies on a clustering of all low-energy
structures evaluated with DFT, with the lowest energy member of each
cluster making up the population. 2) the very final relaxations
in well-sampled basins of the energy landscape, "the final exploitation steps",
are now performed as continued relaxation paths within the first-principles
method, to allow for arbitrarily fine relaxations of the best
structures, independently of the predictive resolution of the
surrogate model. The versatility of the GOFEE method is demonstrated by
applying it to identify the low-energy structures of gas-phase
fullerene-type 24-atom carbon clusters and of dome-shaped 18-atom
carbon clusters supported on Ir(111).
\end{abstract}

\maketitle

\section*{Introduction}


The atomic scale understanding of material properties is a fundamental goal of modern computational chemistry and material science. As material properties to a large extent are governed by the lowest energy atomic structure, the efficient determination of such optimal structure is an important problem. 
The problem is however a difficult one, due to the vastness of the configurational space of even small sized systems, a result of the exponential scaling of the number of metastable structures with the number of atoms in a system \cite{random_search,EA:oganov}.
For simple systems, the use of domain knowledge is often enough to
identify the correct structure. The literature however also contains
many examples, where this approach fails \cite{LEA}, and for the far
majority of systems, that are more complex and less studied, one has
to thoroughly explore the configurational space in order to identify
the optimal structure.
This is commonly achieved using automated and unbiased search strategies such as random search \cite{random_search}, basin and minima hopping \cite{basin_hopping,minima_hopping}, particle swarm optimization \cite{particle_swarm:wang,particle_swarm:CALYPSO}, evolutionary algorithms \cite{EA:johnston,EA:oganov,EA:hammer,EA:GATOR}, etc. which have been successfully applied to an array of different systems including surface reconstructions \cite{reconstruction:TiO2,reconstruction:Si114,reconstruction:KTaO3,SnO2}, grain boundaries \cite{grain_bnd:zande,grain_bnd:xiaowu}, binary compounds \cite{binary:SiO2,binary:review}, isolated \cite{cluster_isolated:RuPt,cluster:isolated:PdCo,cluster_isolated:PdIr} and supported \cite{clusters:alexandrova1,clusters:alexandrova2,LEA,clusters:ZnO} nano-particles, solids \cite{solids:review2019}, etc.

Most often, atomic-scale materials science projects rely on
computationally expensive first-principles methods such as density
functional theory (DFT), van der Waals-enabled DFT, or even more
advanced quantum chemical methods. When using conventional global
optimization methods in conjunction with such energy expressions
their performance becomes limited due to the computational cost of
the many energy and force evaluations required to sufficiently explore the configurational space.
As a potential solution, solving the global optimization problem in
model potentials has proven a capable and less expensive alternative
to solving it with the full energy expression, which we for the sake
of simplicity within this work will refer to as "DFT" or "target potential".
The model potentials have been based on database-trained machine
learning methods, such as kernel based methods \cite{GAP,FF:GDML}
combined with robust representations of atomistic structures
\cite{parrinello_behler,oganov_valle,SOAP}, methods based on
body-ordered energy decompositions \cite{MTP,FF:ortner} and deep
neural networks \cite{parrinello_behler,schnet}.

Despite cheap evaluations once a model is trained, the required quantum mechanical calculated databases represent a considerable computational expense. As an example, for molecular dynamics to be successfully carried out with a machine learned potential, the potential must, depending on the temperature adopted, accurately describe all configurations below a certain energy, which requires a both broad and thorough database. Improvements in data efficiency of such databases have recently been achieved by generating the data using active learning \cite{GAPRSS:boron,GAPRSS:crystal,activeFF:roitberg,activeFF:Zhang,active:shapeev2017,active:ondrej2020,active:smith2020,active:assary2020}, where starting from a small, incomplete database, the model itself, in combination with an acquisition strategy, actively collects all further data, with new data iteratively improving the model throughout the collection process.
Active learning can also be applied in a problem specific context, where further savings in training data are possible, because accurate predictions are especially relevant only for subsets of the full configurational space. This includes configurational problems that are local in nature, such as structure relaxation \cite{local:karsten, local:kastner} and transition state determination \cite{neb:andrew2016, neb:hannes2017, neb:bligaard2019} as well as more ambitious tasks such as molecular dynamics \cite{activeMD:Vita,activeMD:andrew,activeMD:Ohno,activeMD:evgeny,activeMD:Shapeev,activeMD:kresse}, chemical reaction networks \cite{activeReact:Noerskov,activeReact:Stocker} and finally global structure search \cite{SS:alexandrova,active:rinke,active:oguchi,activeSS:calypso,activeSS:shapeev,SS_dftb:maxime,LEA,activeSS:daSilva}, including our recently proposed Global Optimization with First-principles Energy Expressions (GOFEE) structure search method \cite{GOFEE}, which will be further detailed in this work, along with some additional improvements to the population and local convergence of structures.
For global structure search an active learning approach can utilize the fact
that accuracy is increasingly important for lower energy structures,
such that higher energy structures can be screened based on only rough
energy predictions. In GOFEE we employ an actively learned surrogate
model, and aid the active learning in several ways. 1) As a prior for
the surrogate model, we emulate the repulsive interatomic potential,
generally present in atomistic structures, to aid the screening of
irrelevant, high energy structures. 2) in the surrogate model we apply a kernel
with two separate length scales, a longer one responsible for rough energy extrapolation, and a shorter to improve short scale resolution of the model. 3) In each search iteration the single most promising structure is selected, from a number of stochastically generated candidates, by a surrogate model based acquisition function. This acquisition function favors low energies and large predicted uncertainties, naturally supplying an explorative incentive to the search strategy.

The paper is outlined as follows. First a detailed description of the
GOFEE search method is given, including details of the introduced improvements in the population and local convergence capabilities of the method. Throughout, the effect of key elements of the method is illustrated on searches for the reconstructed surface of $\text{TiO}_2(001)\text{-}(1\times4)$. Finally the model is applied to isolated and Ir(111) supported carbon clusters, and we report new low energy structures for the supported system.



\section*{Method}
\begin{figure}
	\centering
	\includegraphics[width=0.9\linewidth]{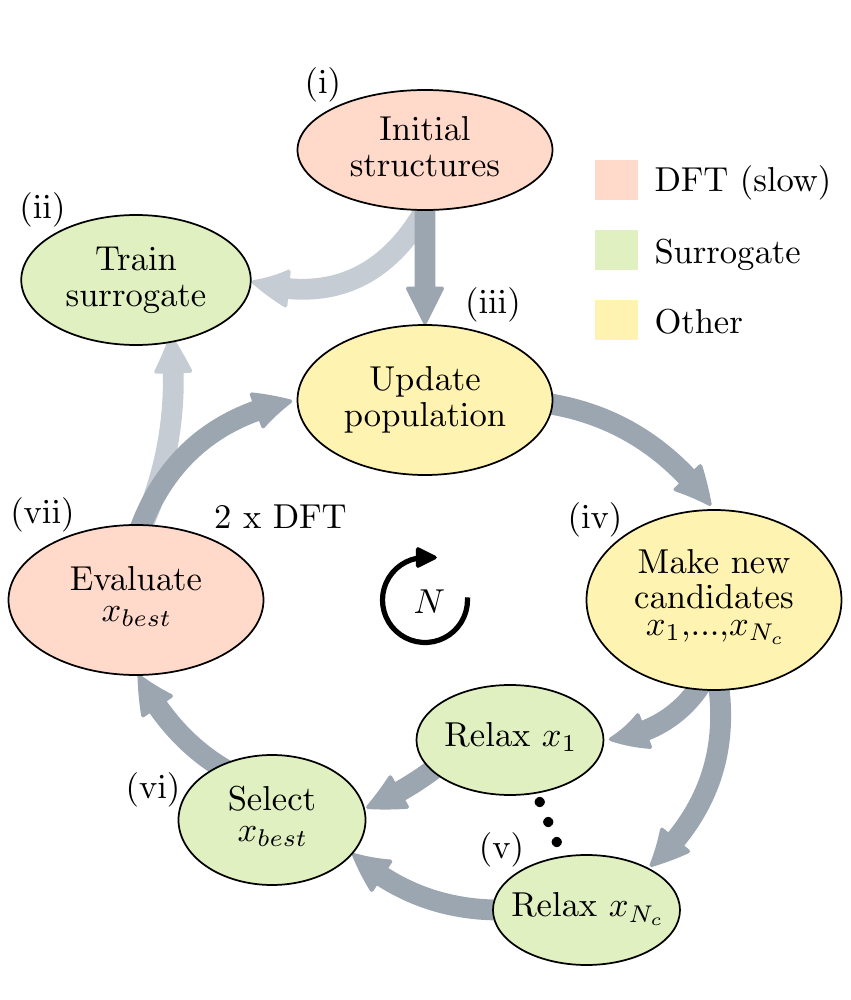}
	\caption{Workflow in GOFEE. (i) Random initial structures are
          generated, evaluated with DFT and added to a structural
          database. (ii) The surrogate model is trained based on the
          database (iii) The population is initialized/updated. (iv) A
          number, $N_c$, of new candidate structures are generated by applying stochastic changes to the population. (v) All $N_c$ candidates are relaxed in the acquisition function. (vi) The single most promising candidate according to the acquisition function is selected. (vii) A single evaluation, using the target energy expression, is performed for this structure, along with a second evaluation for the structure perturbed slightly along the force. The search is carried out by adding these two newly evaluated structures to the training database and repeating steps (ii)-(vii) $N$ times in total.}
	\label{fig:method}
\end{figure}

The GOFEE search method combines the evolutionary search strategy with a computationally inexpensive, actively learned surrogate model of the energy landscape, which can be used to carry out significantly more structure queries, than would be possible with the target potential. A much smaller number of evaluations is however carried out using the target potential on the structures deemed most promising by the surrogate model. These are in turn used as training data to further improve the surrogate model.
A flowchart of the GOFEE search scheme is shown in \cref{fig:method}. The key steps can be summarized as:
\begin{enumerate}[(i)]
	\item Generate and evaluate a number of random structures used to initialize the population and surrogate model.
	\item Train surrogate model based on all DFT evaluated structures accumulated up until this point in the search.
	\item Update population based on all DFT evaluated structures.
	\item Generate new candidates by applying stochastic rattle and permutation operations on the structures currently in the population.
	\item Relax all new candidates using the surrogate model.
	\item Select "most promising" candidate structure. This is determined by the surrogate model.
	\item Evaluate the selected structure using the DFT potential.
\end{enumerate}

\begin{figure}
	\centering
	\includegraphics[width=0.8\linewidth]{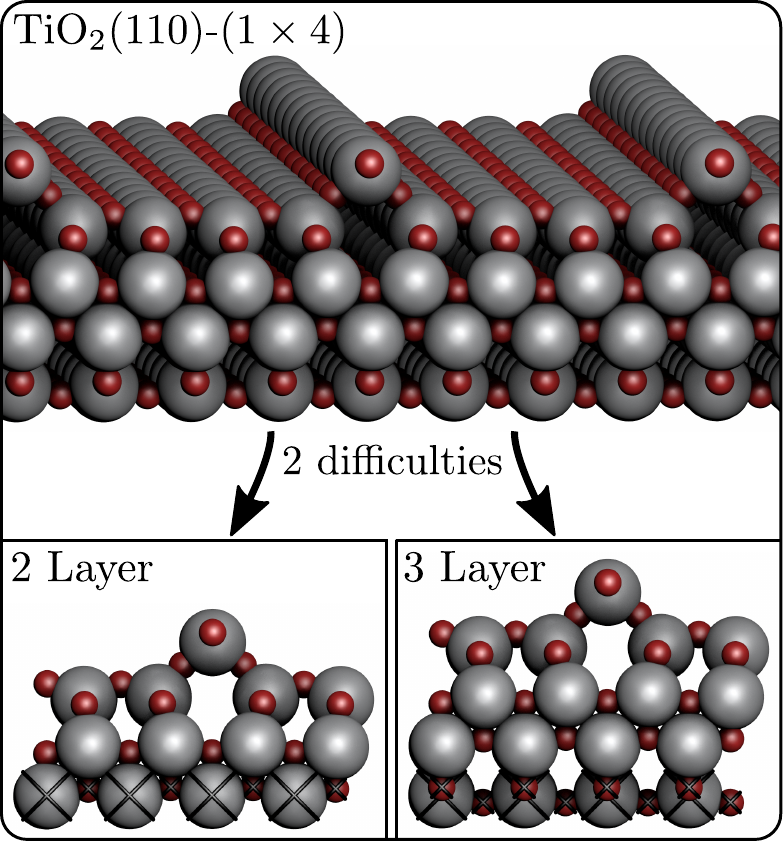}
	\caption{The anatase $\text{TiO}_2(001)\text{-}(1\times4)$ surface reconstruction, featuring rows of protruding titanium atoms. In the structure search context we consider two versions of the problem, having either 2 or 3 atomic layers optimized on top of a fixed bulk layer. One unit cell of the global minimum for these problems, labeled the 2-layer and 3-layer $\text{TiO}_2(001)\text{-}(1\times4)$ problems respectively, are shown in the figure. For the 2 layer problem 27 atoms are optimized, whereas 39 atoms are optimized in the 3 layer problem.}
	\label{fig:tio2_structure}
\end{figure}

The generation of initial structures in the first step (i) is done by randomly placing atoms within a predefined box and requiring that no bond lengths are shorter than $0.7d_{cov,ij}$, with $d_{cov,ij}$ being the sum of the covalent radii of the involved atoms $i$ and $j$. In addition each atom is required to have its nearest neighbor within $1.4d_{cov,ij}$, to avoid isolated atoms.  After this step, the search is carried out by repeating steps (ii) through (vii).
In the following we will discuss in more detail all the above outlined
elements of the search and assess the importance of specific choices that have been made in regards to each of them.
To this end, the effect on the search performance is quantified by
statistics based on multiple independent searches to find the
structure of the anatase $\text{TiO}_2(001)\text{-}(1\times4)$ surface
reconstruction \cite{reconstruction:TiO2}, shown in
\cref{fig:tio2_structure}. As also illustrated in the figure, we will
use two different difficulties of this problem, by requiring the
search to find the correct positions of either the uppermost 2 or 3
layers. We will label these the 2-layer and 3-layer
$\text{TiO}_2(001)\text{-}(1\times4)$ problems, respectively. For
computational convenience, all $\text{TiO}_2$ studies were carried out
using density functional tight-binding (DFTB) theory calculations with
the parameters from Ref.\  \onlinecite{DFTB:TiO2}. Using DFTB as opposed to DFT for these systems does not introduce any significant difference, as the two potentials share the same global minimum and are of comparable difficulty from a structure search perspective.

Such statistical investigations are carried out on multiple occasions as we move on to the discussion of each component in the search, starting with the surrogate model.

\subsection*{(ii) The surrogate model}

The GOFEE search method relies heavily on a computationally inexpensive surrogate model of the energy landscape, to reduce the number of expensive DFT evaluations required to carry out a search.
For this purpose we adopt a Bayesian approach, to have convenient access to prediction uncertainties and specifically choose to use a Gaussian Process Regression (GPR) model, as it is very adequate at learning continuous functions and behaves well even with little training data. This is especially the data-condition in the beginning of a search, where only a small number of structures have yet been evaluated. The data is a set of observed atomic configurations $X=(\mb{x}_1,\mb{x}_2,\dots,\mb{x}_N)^T$ and their energies $\mb{E}=(E_1,E_2,\dots,E_N)^T$. 
To accommodate learning, it is crucial to take advantage of the basic symmetries of the Hamiltonian. This is achieved by letting $\mb{x}_i$ be a suitable representation of the $i$'th configuration. 
In the GOFEE method we follow the approach of Oganov and Valle \cite{oganov_valle} and use, for the representation, a Gaussian smeared distribution of interatomic distances and extend it to interatomic angles as well. The full representation is thus made by binning and concatenating the following distribution functions for all combinations of atomic species, represented by placeholders A, B and C

\begin{align}
F_{A,B}(r) \propto \sum_{A_i,B_j}\frac{1}{r_{ij}^2}\exp\left(-\frac{ (r-r_{ij})^2 }{2l_r^2}\right),
\end{align}
\begin{align}
F_{A,B,C}(\theta) \propto \sum_{A_i,B_j,C_k}f_c(r_{ij})f_c(r_{ik})\exp\left(-\frac{ (\theta-\theta_{ijk})^2 }{2l_\theta^2}\right),
\end{align}
where $f_c$ is the smooth cutoff function
\begin{align}
f_c(r) = 1+\gamma\left(\frac{r}{R_\theta}\right)^{\gamma+1} - (\gamma+1)\left(\frac{r}{R_\theta}\right)^{\gamma}.
\end{align}

For the cutoff function a sharpness of $\gamma=2$ is used. The radial and angular cutoff radii adopted are $R_r=\SI{6}{\angstrom}$ and $R_\theta=\SI{4}{\angstrom}$ respectively and the widths of the Gaussians used for smearing is $l_r=\SI{0.2}{\angstrom}$ and $l_\theta=\SI{0.2}{\radian}$. Finally 30 bins are used to discretize each of the radial and angular distribution functions.

\begin{figure*}
	\centering
	\includegraphics[width=0.9\linewidth]{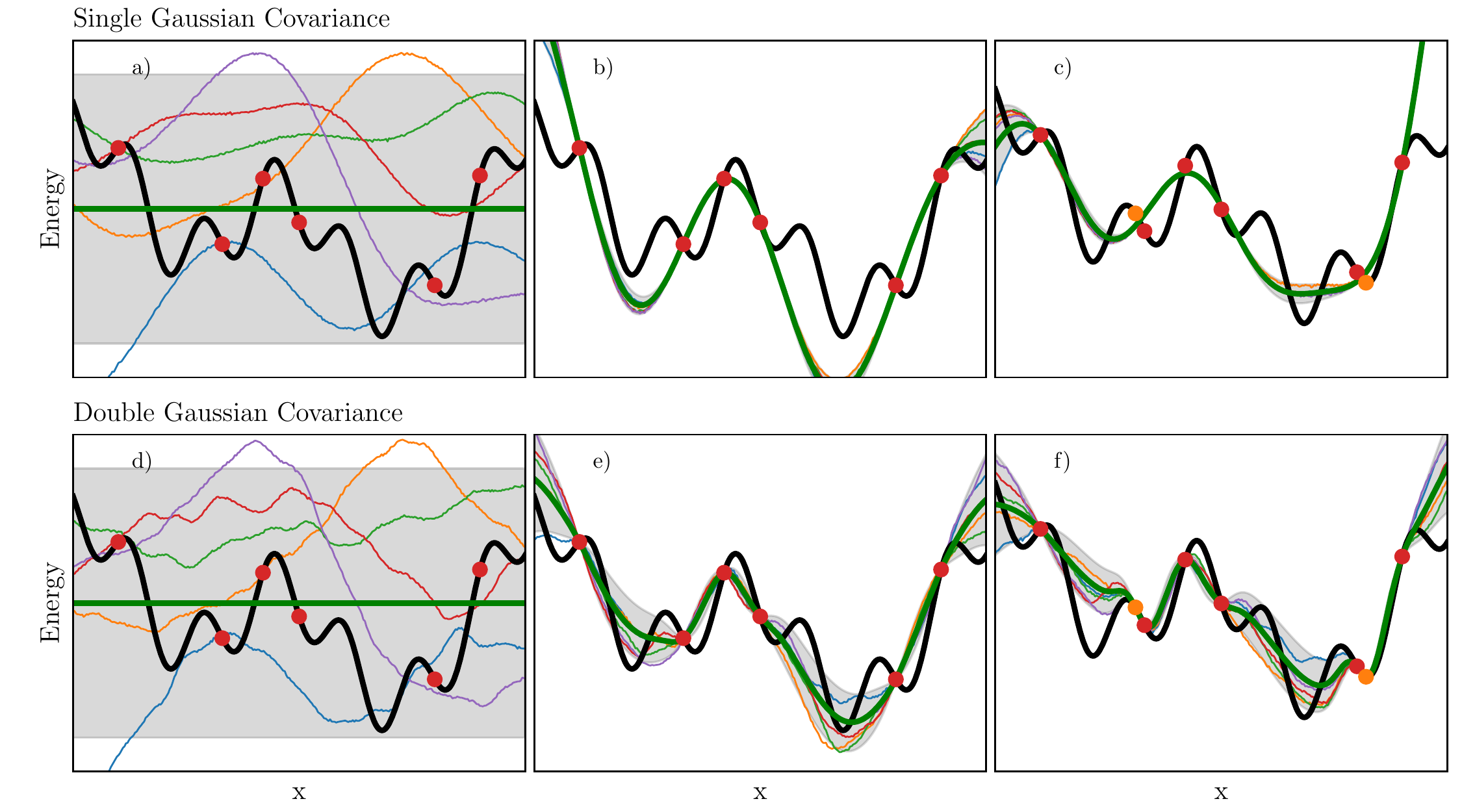}
	\caption{One dimensional GPR example, comparing the normal
          Gaussian kernel (a,b,c) with the double Gaussian kernel
          (d,e,f) used in GOFEE. The target function are represented
          by the thick green line, the GPR model by the black line and
          the training data by red dots. The thin lines represent
          functions sampled from the model distributions. a and d)
          Prior to training, both models just predict the mean of the
          data, however functions sampled for the two kernels are
          visually dissimilar, with the double Gaussian kernel
          resulting in short scale wiggles on top of the long scale
          variations. b) and e) For the trained models, the single
          Gaussian model tends to overshoot the target, and
          underestimate short scale uncertainties
          Even with only a weight of $\beta=0.01$, the short scale Gaussian in the double Gaussian model remedies these flaws to a large extent.
		c) and f) Adding further data, as orange dots, requiring even more short scale variation, only reinforces this point.}
	\label{fig:covariance}
\end{figure*}

Representing atomic configurations with feature vectors $\mb{x}$ in this form, the GPR model is then tasked with inferring a distribution over functions, $p(E_{sur}|X,\mb{E})$, that is consistent with the training data $(X, \mb{E})$.
As this distribution, by definition of a Gaussian Process, is assumed
normal, its mean and standard deviation respectively can be used to
predict both the energy, $E_{sur}(\mb{x})$, and the expected
uncertainty, $\sigma_{sur}(\mb{x})$, on the energy prediction, for a
new atomic configuration.  The GPR model is defined by its prior mean
function $\mu(\mb{x})$ and covariance function $k(\mb{x}_i,\mb{x}_j)$,
from which the prior distribution, i.e. the untrained model, is
derived. Given these, the posterior distribution, i.e. the trained
model, is determined by conditioning the prior distribution on the available training data. The resulting expressions, for energy and uncertainty prediction of a new structure $\mb{x}_*$, are \cite{GP:Rasmussen}
\begin{align}
E_{sur}(\mb{x_*}) &= \mb{k}_*^T(K+\sigma_n^2 I)^{-1}(\mb{E}-\mu(\mb{x}))+\mu(\mb{x}), \\[4pt]
\sigma_{sur}(\mb{x_*})^2 &= k(\mb{x}_*,\mb{x}_*) - \mb{k}_*^T(K+\sigma_n^2 I)^{-1}\mb{k}_*,
\end{align}
where $K = k(X,X)$ and $\mb{k}_* = k(X,\mb{x}_*)$, and the target function is assumed noisy with uncertainty $\sigma_n^2=5*10^{-2} \SI{}{\electronvolt}^2$, which acts as regularization.
\Cref{fig:covariance}a-c illustrates a GPR model applied to data sampled from a one dimensional function. Both the prior and posterior distributions are depicted with their mean and standard deviation, along with concrete sample functions from the two distributions.

The GPR model in \cref{fig:covariance}a-c uses the Gaussian covariance
\begin{align}
k(\mb{x},\mb{x}') = \theta_0e^{ -(\mb{x}-\mb{x}')^2/(2\lambda^2) }
\label{eq:covariance_single}
\end{align}
with characteristic length scale $\lambda$ and maximal covariance $\theta_0$. In GOFEE we instead adopt a covariance function consisting of a sum of two Gaussian covariance functions with different length scales, which we dub the double Gaussian covariance.
\begin{align}
k(\mb{x},\mb{x}') = \theta_0 [ (1-\beta)e^{ -(\mb{x}-\mb{x}')^2/(2\lambda_1^2) } + \beta e^{ -(\mb{x}-\mb{x}')^2/(2\lambda_2^2) } ]
\label{eq:covariance}
\end{align}
with characteristic length scales $\lambda_1$ and $\lambda_2$ respectively, maximal covariance $\theta_0$ and weights given by $\beta=0.01$. As is common for GP models, all non-fixed hyperparameters are automatically selected by maximizing the marginal log-likelihood of the parameters given the observed data. In GOFEE we use multi restart gradient decent to carry out the optimization.
The present covariance function is chosen because the optimized length scale of a single Gaussian covariance tends to be significantly larger than the feature space distance between neighboring local minima and comparable to the extent of the training data in the feature space. This limits the resolution of the resulting model, which in addition tends to be overconfident in its predictions. Adding the second Gaussian with a small weight and shorter length scale than the first, partly remedies this problem.

The effect of this choice of covariance is exemplified in
\cref{fig:covariance}, which in addition to, \cref{fig:covariance}a-c,
showing the model resulting from using the normal Gaussian covariance,
\cref{eq:covariance_single}, also shows, in \cref{fig:covariance}d-f,
the result of using the double Gaussian covariance,
\cref{eq:covariance}. The compared covariance functions are identical in all
but the shorter length scale Gaussian, which has only a weight
$\beta=0.01$, relative to the long length scale Gaussian. Even this
small addition of a shorter length scale is remarkably visible,
already in the functions sampled from the priors,
\cref{fig:covariance}a and d. Moving on to the trained models, the
addition of the shorter length scale, results in models, which are
better at accommodating short scale changes in the target function,
and which do not suffer from the underestimation of uncertainties on
the short scale, to the extent as when using the normal Gaussian
covariance. We have recently used the double Gaussian covariance
in the context of structure optimization with image recognition and
reinforcement learning\cite{Mortensen_2020}, where it proved highly
efficient in speeding up the searches.

\begin{figure}
	\centering
	\includegraphics[width=1\linewidth]{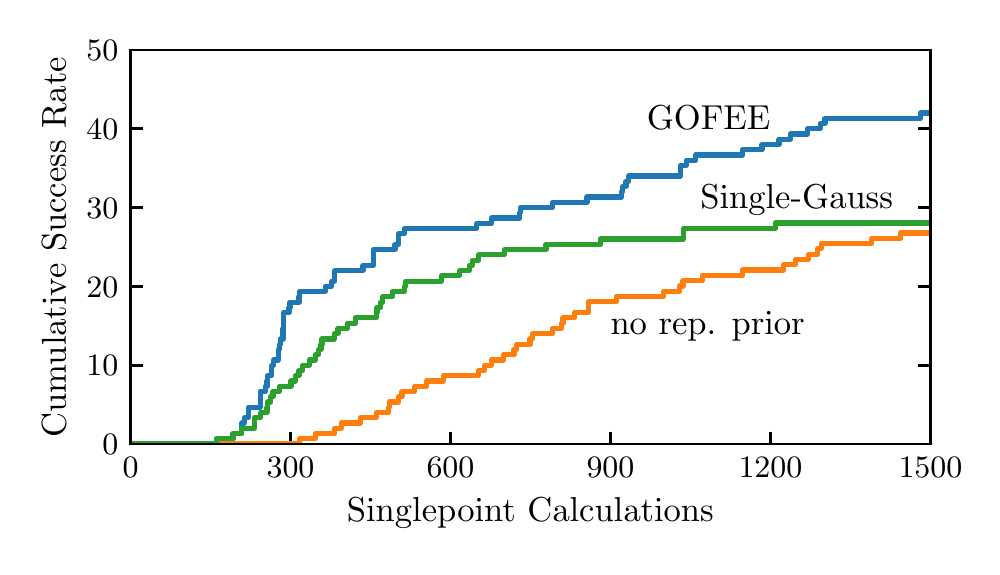}
	\caption{Benchmark of GOFEE, on the 3-layer
		$\text{TiO}_2(001)\text{-}(1\times4)$ problem. Blue curve: GOFEE is used as presented in this paper, with
		dual-point, the double Gaussian kernel, the same
		$\kappa_{relax}=\kappa=2$ used for relaxation and candidate
		selection and with the repulsive prior. Green curve: uses the normal Gaussian covariance. Orange curve: Omitting the repulsive prior mean function.}
	\label{fig:success_benchmark_cov_prior}
\end{figure}

Because the acquisition function, central to GOFEE, depends on the predicted uncertainties, the underestimation of these causes a less lively search, more prone to stagnation, as the uncertainties contribute the exploitative incentive of the acquisition function. This is reflected in the results in \cref{fig:success_benchmark_cov_prior}, showing that the searches adopting the normal Gaussian kernel are more prone to get stuck, compared to the double Gaussian kernel.

Besides the covariance function, the second component to the prior distribution of the GPR model, is the prior mean function, $\mu(\mb{x})$. For this a common and simple choice is to use a constant value equal to the mean, $\bar{\mb{E}}$, of the training data. It is however useful to keep in mind that the model effectively only needs to learn the difference between the prior mean function and the target function. It can therefore be useful to include general features of the target function, here the total energy, into $\mu(\mb{x})$. In GOFEE we add to the data mean a conservatively chosen, repulsive interatomic potential, such that $\mu(\mb{x}) = \bar{\mb{E}} + \sum_{ij}(0.7\cdot r_{CD,ij}/r_{ij})^{12}$, where $r_{CD,ij}$ is the sum of the covalent radii of the $i$'th and $j$'th atoms and $r_{ij}$ is the distance between them. This is used with the main purpose of not sampling the very high energy structures resulting from too short bonds, as such data can significantly impact the energy scale of the regression problem and negatively affect the resulting model.

Without the repulsive prior, the search naturally, and especially early on, tends to spend more resources sampling structures with unreasonably short bonds, which typically have high energies. As a derived effect the surrogate model also has to accommodate a significant amount of high energy structures, which to some extent compromises the prediction accuracy on the low energy structures, which are relevant to the search. This derived effect is largest early on, where the high energy structures make up a larger proportion of the data.
As shown in \cref{fig:success_benchmark_cov_prior}, neglecting the repulsive prior results in a less effective search strategy, with the difference being especially apparent early on in the searches.

Sticking with the order defined in the flowchart, we will now halt the discussion of the surrogate model for a while, in favor of a description of how the population is maintained and then used as the basis for generating new candidates. We will then return to the surrogate model, when it is used for relaxation of these new candidates.

\subsection*{(iii) Population and new candidates}
\begin{figure*}
	\centering
	\includegraphics[width=0.9\linewidth]{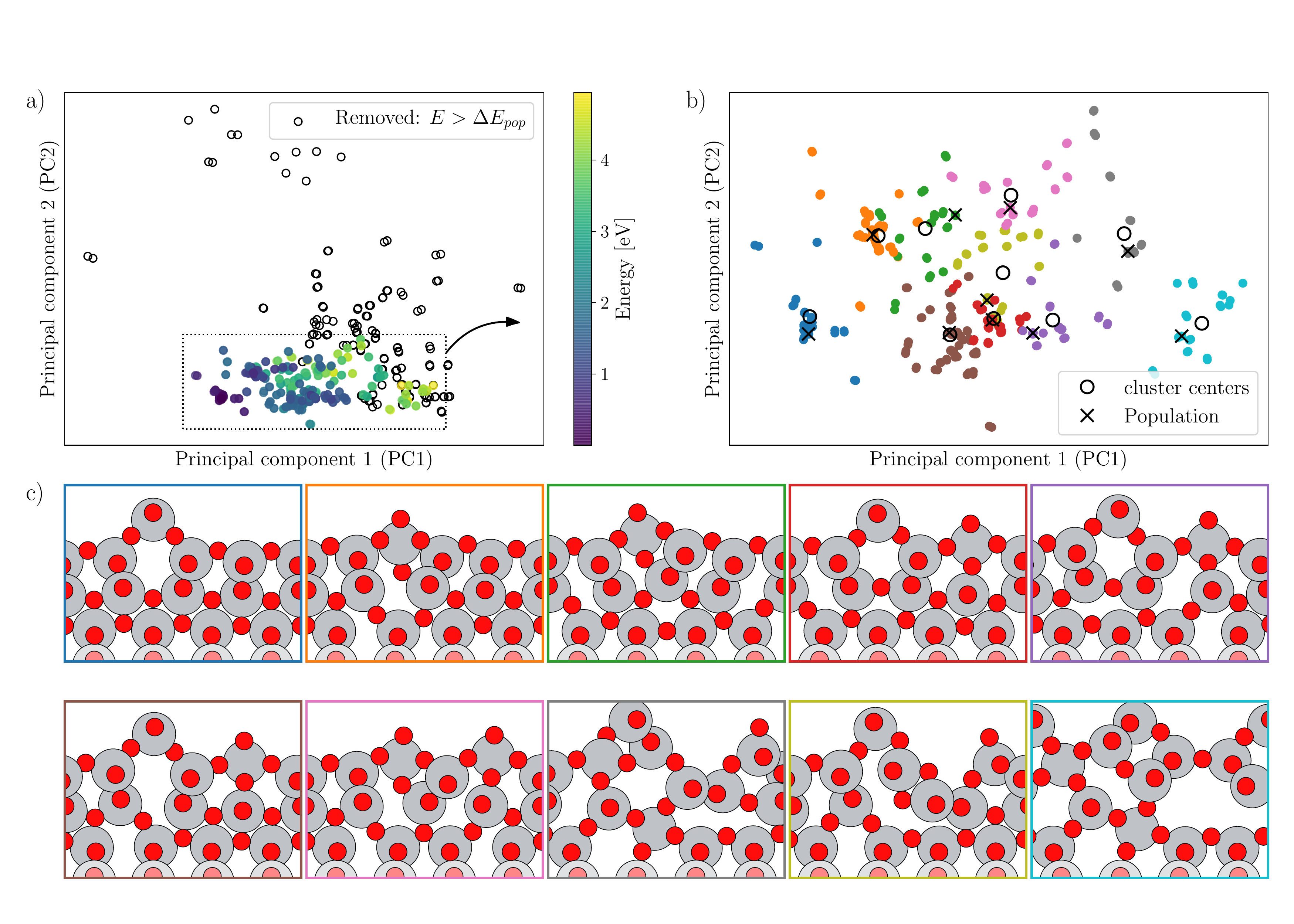}
	\caption{The population scheme, visualized with principal
          component analysis (PCA), with the first two principal
          components capturing $83\%$ of the variance. a) First, all
          evaluated structures, with an energy above $\Delta E_{pop}$
          of the best structure found so far, are ignored, as
          represented by black circles. b) Secondly, The remaining
          structures are clustered in the feature space using the
          k-means++ algorithm. The resulting clusters are colored, and
          cluster centers drawn as black circles. Thirdly, the
          population is created by selecting the lowest energy
          structure in each cluster, drawn as black crosses. c) The population resulting from applying the scheme to this dataset, consisting of the 500 first structures of a GOFEE search. As the population structures can lie on cluster edges, some could in principle be very similar. The diversity of the population is however not found to suffer greatly from this effect, as exemplified in c).
		The adequacy of the PCA visualization is also visible in the fairly reasonable positions of the cluster centers, and only slight smearing of the cluster borders.}
	\label{fig:population}
\end{figure*}
The population is an important element for any evolution based search strategy, as it effectively controls what new candidate structures are likely to be created. To avoid premature convergence of the search, while still progressing it, the population must be both diverse and prioritize low energy structures. In the present work we propose a new approach, which achieves a diverse population through clustering the previously evaluated structures.
The scheme is sketched in \cref{fig:population} for real GOFEE data using principal component analysis (PCA). As depicted in \cref{fig:population}a the scheme involves first selecting all structures within an energy span $\Delta E_{pop}$ of the best structure found so far in the search. $\Delta E_{pop}=\SI{5}{\electronvolt}$ is used in this work. This prioritizes low energy structures and prohibit the, in this context undesired, high energy structures, also present in the structure database, from affecting the clusters. 
Secondly, as shown in \cref{fig:population}b the selected structures are clustered in the feature-space using the distance based clustering method, k-means. Finally, the population is built by selecting the lowest energy structure in each cluster. \Cref{fig:population}c depicts the population resulting from applying this population scheme.
This represents an improvement compared to Ref.\ \onlinecite{GOFEE},
where the population was maintained by applying a maximum covariance
threshold of $k_{max}=\SI{0.9995}{\electronvolt \squared}$ to the
lowest energy structures in the database.

The presently proposed means of using clustering in setting up the
population for GOFEE differs from our previous use of clustering in
conjunction with an algorithm approach. In the work of J{\o}rgensen
\textit{et al.}\cite{jorgensen_2017} new members for the population
were chosen based on energy (fitness) and dissimilarity to lower
energy members of the population. Selection of population members
to become parents was made with a scheme favoring
population members that were outliers when clustering the entire set
of calculated structures. In the present work, we construct the population
directly via the clustering, have a uniform parent selection scheme for
candidate production, and choose the optimal candidate for DFT
evaluation via Bayesian statistics, as detailed below. 

\subsection*{(iv) Generating new candidates}
After updating the population this way, $N_c$ new candidate structures are generated by applying stochastic changes to the structures in the population. At present, these include simple operations such as shifting or permuting atoms.
It is worth pointing out, that generating multiple new candidates is only beneficial because we have the computationally inexpensive surrogate model, which can be used to relax and compare them all before it is finally applied to select a single one, to be evaluated with DFT.
Since population members were relaxed in the surrogate model as it
appeared based on the DFT data available at the time of the creation of
the population members, it serves a purpose to reoptimize the
population members with newer versions of the surrogate model. We
hence occasionally add to the $N_c$ generated candidates, the current
population. Specifically, it is done in every third search iteration. This is useful to allow for further optimization of the best structures found so far, but because only a single one of the candidates is selected for DFT evaluation, it comes at the expense of less resources spent on exploration.
This particular compromise ensures that at least two thirds of the resources are spent on some degree of exploration.

We will now turn back to the surrogate model, and discussed how it is used to define an acquisition function for relaxing the new candidates and selecting the most promising one among them.

\subsection*{Relax new candidates}
\begin{figure}
	\centering
	\includegraphics[width=0.9\linewidth]{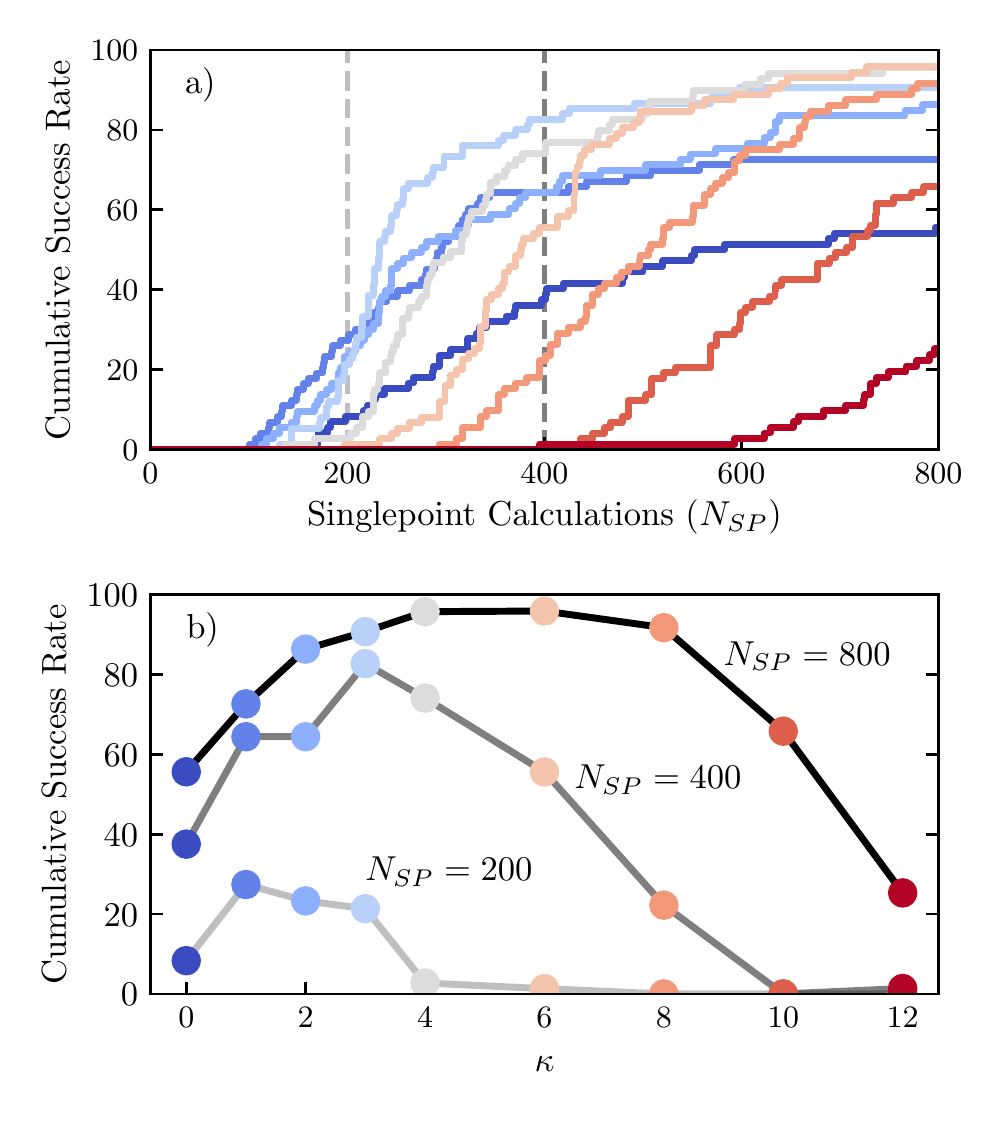}
	\caption{The effect $\kappa$ on the performance on searches on the 2-layer $\text{TiO}_2(001)\text{-}(1\times4)$ system. a) success curves, each based on 75 independent searches, for different values of $\kappa$, with values increasing when going from blue to red. b) Snapshots of the success rates for each curve, in a), $N_{SP}=200$, $400$ and $800$ single-point evaluations into the search.}
	\label{fig:successkappa}
\end{figure}

At this point we take advantage of the surrogate model, which is continuously trained throughout the search, as new data becomes available. However the predictions of the surrogate model are in many regions of the search space lacking due to sparsity of the training data. This is especially true early on in the search, where only very little training data has yet been accumulated. To account for this, it is useful to also rely on the predicted uncertainty, when using the surrogate model to guide the search. This is done by introducing an acquisition function \cite{acquisition}, $f(\mb{x})$, which combines the predicted mean, $E_{sur}(\mb{x})$, and standard deviation, $\sigma_{sur}(\mb{x})$, of the surrogate model. In this work we have used the lower confidence bound \cite{phoenics,explore:mathias2018} of the model predictions,
\begin{align}
f(\mb{x})=E_{sur}(\mb{x}) - \kappa \cdot \sigma_{sur}(\mb{x}) ,
\label{eq:aquisition}
\end{align}
 due to its simplicity. Here $\kappa$ effectively controls the emphasis on exploration in the search. The acquisition function is used both to relax all new candidates and to select the one among them for DFT evaluation.
 \Cref{fig:successkappa}a shows the success curves, for finding the global minimum structure of the 2-layer $\text{TiO}_2(001)\text{-}(1\times4)$ system. The curves represent the expected performance from running GOFEE with different values of $\kappa$. Each curve is produced from 75 independent GOFEE restarts. \Cref{fig:successkappa}b shows snapshots of the corresponding success rates at different points in the search.
\Cref{fig:successkappa}b makes it clear that the pronounced tendency
to exploit, associated with small values of $\kappa$, results in some
searches finding the GM very quickly. For instance, $\sim$30{\%} of searches with
$\kappa=1$ do find the GM within 200 evaluations. Using $\kappa=1$, however, results in a relatively
small long term success rate of $\sim$70{\%} within 800 evaluations, when
compared to the more explorative searches, such as 
$\kappa=4-8$, achieving success rates of 90{\%}  within 800 evaluations. Even larger values of $\kappa$ are expected to reach similar or higher success rates, if the searches were allowed even more evaluations.
Based on this, a choice in the range $\kappa=2-4$ is suspected to strike a reasonable balance between exploration and exploitation. In this work we have used $\kappa=2$ for all searches, but note that if the computational budget allows, adopting a slightly larger value of $\kappa$ in searches run for more iterations, will benefit the thoroughness of the search.

\begin{figure}
	\centering
	\includegraphics[width=1\linewidth]{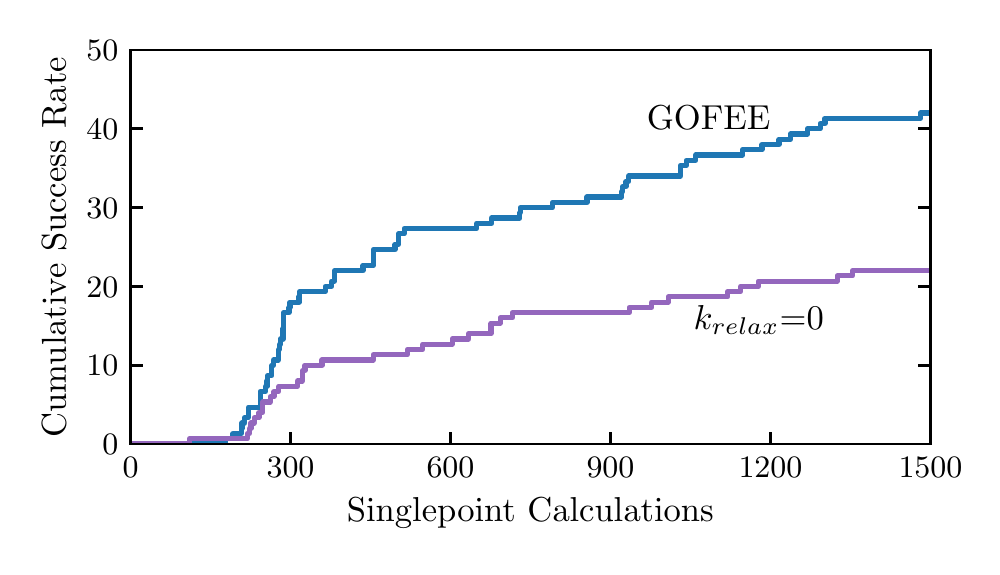}
	\caption{Benchmark of GOFEE, on the 3-layer
		$\text{TiO}_2(001)\text{-}(1\times4)$ problem. Blue curve: GOFEE as presented in this paper. Purple curve: Relaxations are carried out with $\kappa=0$ in the acquisition function.}
	\label{fig:success_benchmark_kappa_relax}
\end{figure}

A valid alternative to using the acquisition function for relaxation
and selection of candidates, is to only use it for selection, and
instead use only the posterior mean of the surrogate model for
relaxation. The posterior mean is the models best guess of the target
potential, and therefore also of the locations of local minima, which ultimately are the structures we are after. This
approach does however perform significantly worse than the adopted
approach, when compared on the 3-layer
$\text{TiO}_2(001)\text{-}(1\times4)$ system, as shown in
\cref{fig:success_benchmark_kappa_relax}. 
The main reason  is likely, that leaving out the bias towards uncertain structures during the relaxations, naturally results in less diversity in the training database containing the evaluated structures. It is thus the active learning element of the search method which suffers.

\subsection*{Select candidate}
\begin{figure}
	\centering
	\includegraphics[width=1.0\linewidth]{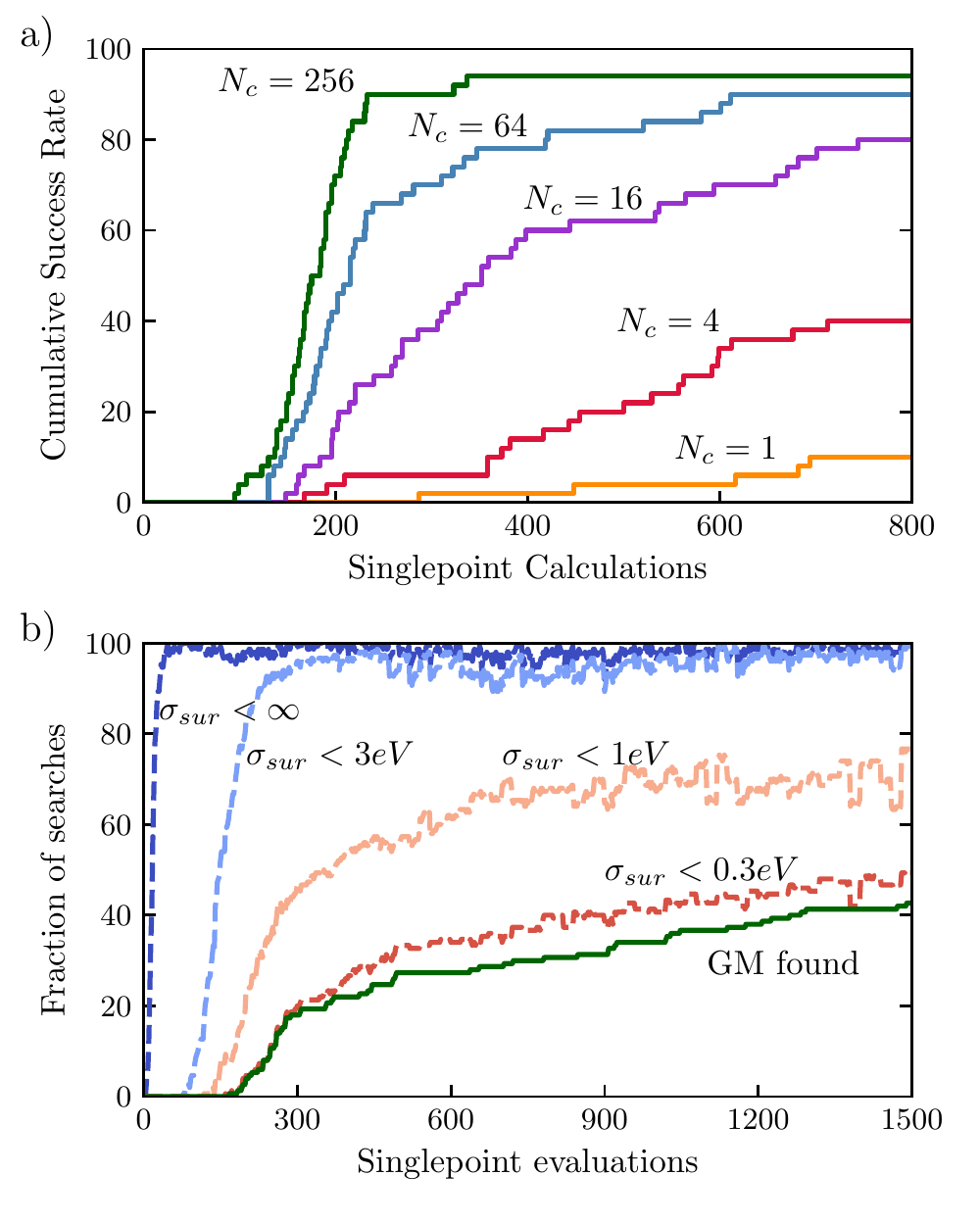}
	\caption{Importance of the new candidate structures. a) Success curves for the 2-layer $\text{TiO}_2(001)\text{-}(1\times4)$ problem, resulting from searches with different numbers, $N_c$, of new candidates generated in each search iteration. The gain in performance, from e.g. quadrupling $N_c$, is remarkable, especially in the range of relatively small $N_c$. b) The full, green line shows the success curve for the 3-layer $\text{TiO}_2(001)\text{-}(1\times4)$ problem. The dashed, dark blue line shows the fraction of the searches, that, based on the acquisition function, would select the global minimum, were it supplied as one of the new candidates.
	The remaining dashed lines only counts this artificial selection of the global minimum successful if the predicted uncertainty, $\sigma_{sur}(\mb{x}_{GM})$, thereon, is less than the various values given in the figure. This supplies information about how certain the model is about the GM, when it is selected over the other candidates.}
	\label{fig:success_candidates}
\end{figure}

Having established the acquisition function for selecting among the new candidates, it is interesting to investigate how efficiently it chooses from among the $N_c$ new candidates, and in particular to what extent it is worth to increase the number of candidates from which the acquisition function chooses.
As long as the acquisition function is better than a random selection scheme, at selecting the candidate best suited at progressing the search, increasing $N_c$ should statistically benefit the search. However in practice, $N_c$ is limited by the required computational time as compared to that of the DFT evaluations. \Cref{fig:success_candidates}a shows the success curves for different values of $N_c$. It illustrates the dramatic importance of generating multiple candidates in each search iteration. When using $N_c=1$ the problem is barely solvable, with only $10\%$ of searches successful, in 800 DFT evaluations, whereas $90\%$ of searches are successful in only little more than 200 evaluations, when using $N_c=256$. This dramatic improvement of the search performance, with the number of new candidates generated in each search iteration, highlights the importance of surrogate models being not only accurate, but also efficient to evaluate. Efforts to reduce the computational cost of evaluating machine learned regression models \cite{fastSOAP,descriptor:bessel,FF:ortner} are therefore highly relevant.

Another way to assess the quality of the surrogate based acquisition function, and whether larger $N_c$ and better operations for generating candidates are worth while, is, to inspect to what extent, the acquisition function would actually select the global minimum, were it supplied alongside the new candidates. As shown in \cref{fig:success_candidates}b, already long before the global minimum is found in any search, and to a remarkable extent, the acquisition function prefers the global minimum over any of the other new candidates. 
This does not tell us, that the model has already identified the global minimum, as there are likely other structures, not among the candidates, that would be selected even over the global minimum, especially in the early stages of the search. It does however tell us that the set of new candidates can be significantly improved, which can be achieved either by increasing $N_c$, by improving the operations used to generate new candidates or by improving the population.

\subsection*{Evaluation}
\begin{figure}
	\centering
	\includegraphics[width=1\linewidth]{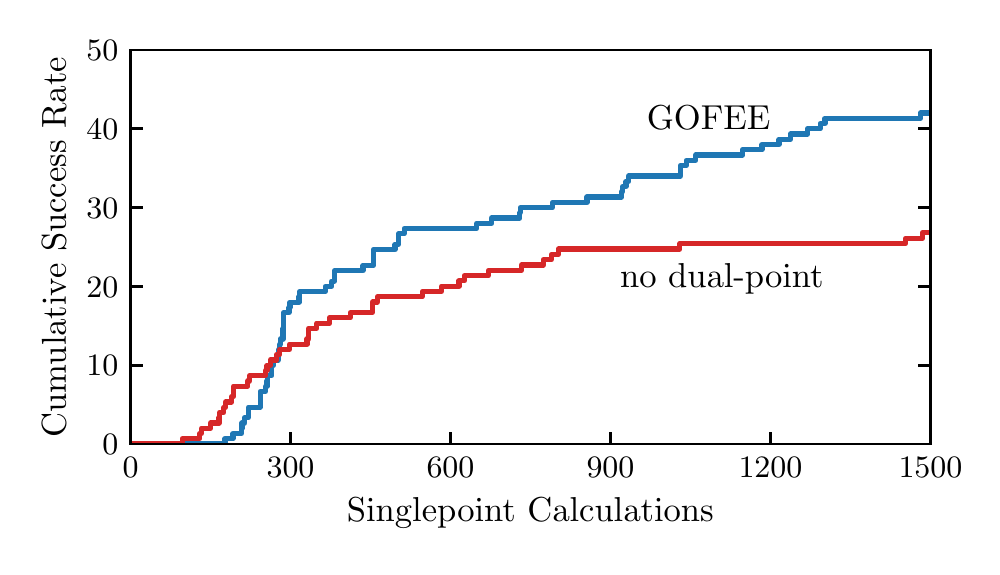}
	\caption{Benchmark of GOFEE, on the 3-layer
		$\text{TiO}_2(001)\text{-}(1\times4)$ problem. Blue curve: GOFEE as presented in this paper. Red curve: The dual-point evaluation is ommitted in the search.}
	\label{fig:success_benchmark_dualpoint}
\end{figure}

As the final step in a search iteration, the selected structure is
evaluated using DFT. This consists of a single-point evaluation of the
structure itself and in addition we make a second evaluation of the
structure perturbed slightly in the direction of the force acting on
each atom. This is done to take advantage of the forces readily
available from DFT codes at little additional cost, when having
already evaluated the energy. As seen from
\cref{fig:success_benchmark_dualpoint},
this is an improvement compared to evaluating only the structure itself, omitting the second evaluation.
The likely rational for this improvement, over actively letting the acquisition function select all new structures, is, that in the present context of structure search, where the objective is to minimize the energy, the direction of the force is the most valuable direction in which to improve the surrogate model.
One can also take advantage of the force information by explicitly including it, when training the GPR model \cite{FF:GDML,local:karsten,local:kastner,neb:hannes2017,neb:bligaard2019}. This is not adopted in GOFEE, although we have explored the approach, the reason being significantly longer training and evaluation times, which makes it unfit for the present setting, despite improving model predictions.

As an addition to Ref.\ \onlinecite{GOFEE}, we here further propose a concurrent evaluation scheme, which is only applied, when the structure selected for evaluation is close to being fully relaxed. It is introduced to improve the capability of GOFEE to locally optimize structures, to arbitrary precision, during the search. This is relevant because we use a global descriptor in the GPR based surrogate model, which inevitably limits the models ability to accurately resolve local minima, and thus fully relax structures.

\begin{figure}
	\centering
	\includegraphics[width=0.8\linewidth]{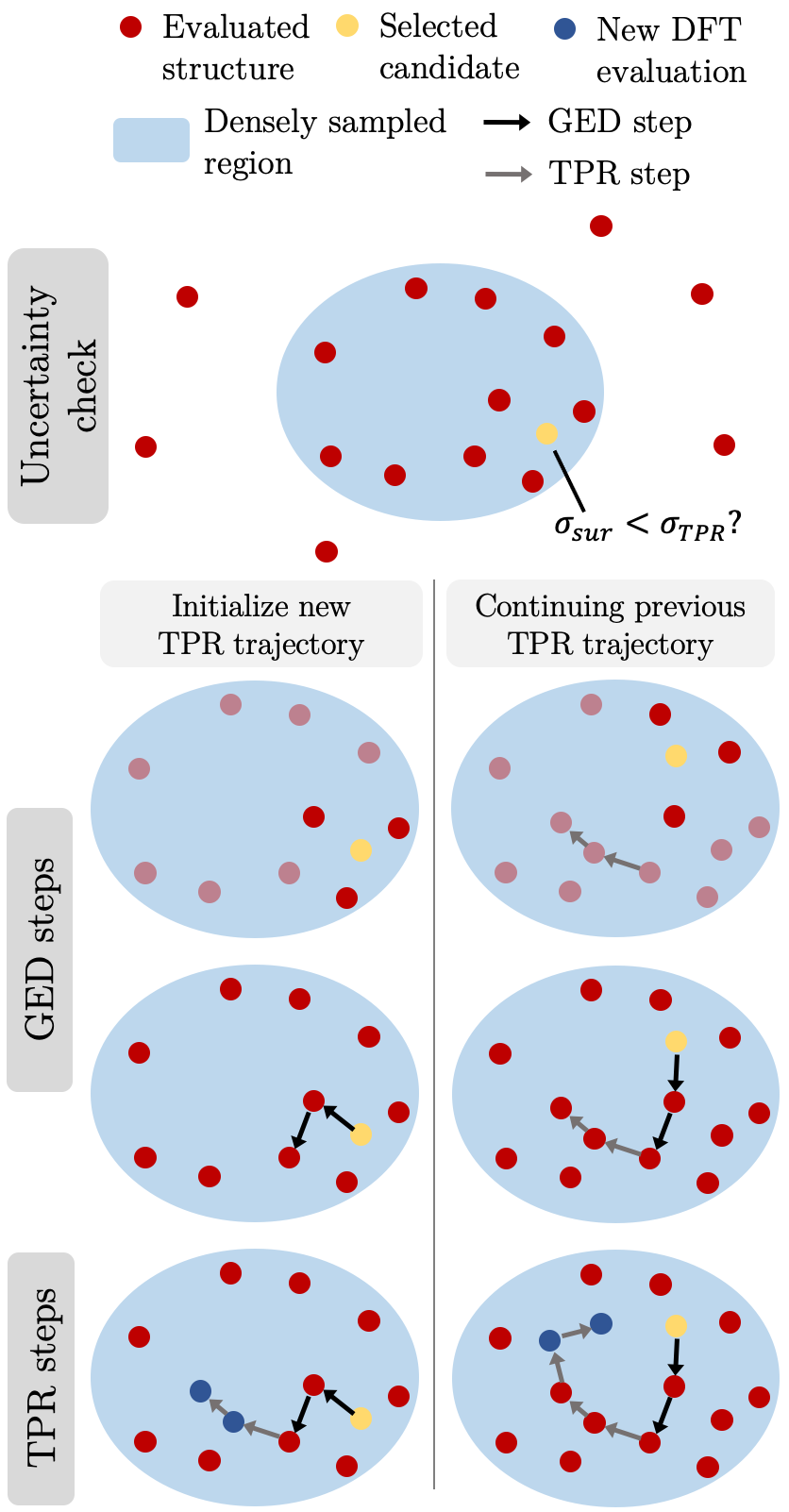}
	\caption{Sketch of the TPR scheme, with $N_{GED}=3$ for
          simplicity. First, the predicted uncertainty of the selected
          structure is compared to the threshold value $\sigma_{TPR}$,
          to determine, whether to continue with the TPR scheme. If
          moving on with the scheme, GED steps are taken, in an
          attempt to reach the lowest energy structure among the
          previously evaluated structures within the basin. Each GED
          step is carried out by moving to the nearest structure,
          among the $N_{GED}$ closest, which has a lower energy than
          the present one. When no lower energy neighbors are found, a
          TPR trajectory is initialized by carrying out two relaxation
          steps in the target potential, from the final
          structure. Alternatively, if the GED steps at some point
          reaches a structure which is part of an ongoing TPR
          trajectory, this relaxation is continued.}
	\label{fig:tpr}
\end{figure}

As a solution, we propose the target potential optimization (TPR)
scheme, shown in \cref{fig:tpr}. Here, we propose to  use the
surrogate model itself to identify cases where it is inefficient to
use for further relaxation. These cases are defined to be whenever the
predicted uncertainty of the structure selected for evaluation is
sufficiently small, which naturally occurs in regions, where a basin
around a local minimum has been densely sampled. Specifically we use the criterion $\sigma_{sur}<\sigma_{TPR}$, with $\sigma_{TPR}=\SI{0.1}{\electronvolt}$.

In cases where this holds true for the uncertainty, rather than
spending two DFT evaluations following the dual-point scheme, we opt
to spent two DFT evaluations on initializing or continuing a TPR
trajectory.
In order to determine from which structure to initialize the relaxation or locate a previously initialized one, we use a greedy energy descent (GED) scheme.
The GED scheme involves moving to the closest structure, in the
feature space, with lower energy than the present one, considering in
each step only the $N_{GED}$ closest structures, among the previously
evaluated ones. This is repeated until none of the $N_{GED}$ closest
structures are lower in energy or until a structure, belonging to an
ongoing TPR trajectory, is reached, in which case this relaxation is progressed further. In this work, we have used $N_{GED}=5$.
This simple scheme gets the far majority of assignments correct, and avoids the need to a characteristic distance in the feature space, which would have to be system dependent. To get all assignments correct, one would, in any case, have to deviate from such purely distance based approaches, to account for the anisotropic nature of energy basins, as seen from the feature space.

\section*{Applications}
To showcase the versatility of the method we finally present
two full-scale DFT structure searches. Specifically, we have applied
the method to gas-phase C$_{24}$ clusters \cite{C24}, and Ir(111)
surface bound C$_{18}$ clusters.

\subsection*{Carbon clusters}
\begin{figure}
	\centering
	\includegraphics[width=0.9\linewidth]{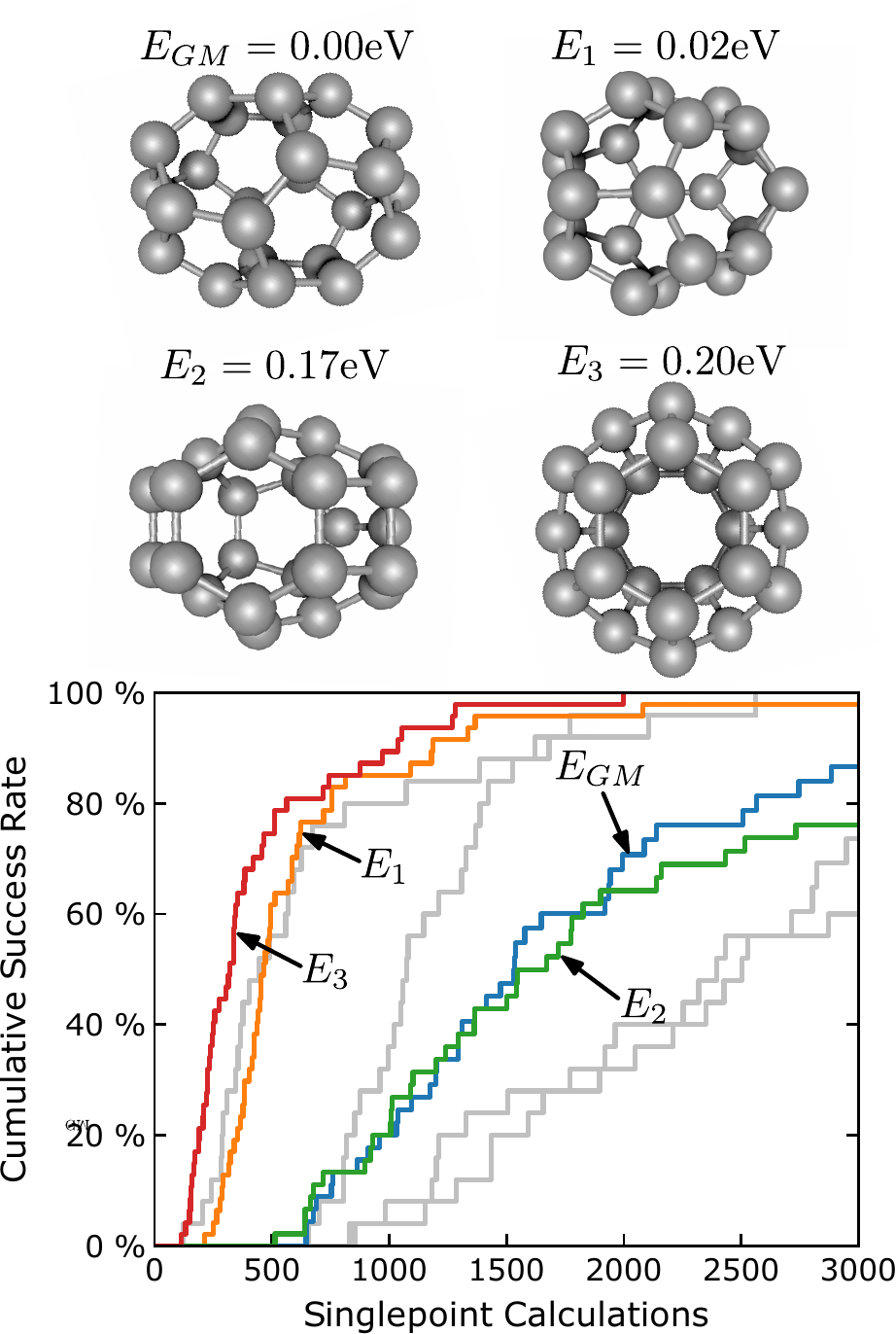}
	\caption{The geometries of the four lowest energy structures of C$_{24}$, along with success curves for finding each structure, based on $50$ independent restarts of GOFEE. The colored success curves represent searches run with the new population and improved local convergence schemes, whereas the gray curves represent searches without these additions. Structures 1 and 3 are found in almost $\SI{100}{\percent}$ of searches after only $1500$ DFT evaluations, whereas the global minimum and structure 2 is found in $\approx\SI{80}{\percent}$ of searches after $3000$ DFT evaluations. These might be more difficult because they both feature two four membered rings, which are typically not energetically favored compared to hexagonal and pentagonal rings. In comparison, structures 1 and 3 contain only one and none, respectively.}
	\label{fig:success_c24}
\end{figure}

In the case of the C$_{24}$ clusters, the potential energy landscape exhibits multiple local minima, which are very close in energy to the global minimum, but which are geometrically quite different, as seen in \cref{fig:success_c24}, which shows the geometry of the four lowest energy structures for the system. The GOFEE search method relies to a very large extent on the surrogate model, trained on the fly, from scratch. One might therefore be concerned about the possibility, that the search will tend to overlook local minima, due to prediction errors in the trained model on structures it have not yet seen. As seen from \cref{fig:success_c24}, which also shows the success curves for finding each of the four structures, this is not the case to any noticeable extent, as almost $\approx\SI{80}{\percent}$ of searches find all four structures within $3000$ DFT evaluations. This is not because the model is accurate to within $\SI{0.02}{\electronvolt}$, on geometries not present in the training data, but because of the prediction uncertainty, present in the acquisition function, which drives the search towards exploring new geometries, which might not have been selected for evaluation, were the acquisition function based on the predicted energy alone.

\subsection*{Carbon clusters on Ir(111)}

In recent years, graphene nanostructures, such as nanoribbons and
nanoislands, have attracted increasing attention, due to the
tunability of their electronic and chemical properties, achieved by
varying the shape and size of such structures
\cite{suppC:dresselhaus2016}. Compared to nanoribbons, graphene
nanoislands have the largest potential for variation, owing to the
larger number of configurational degrees of freedom
\cite{suppC:snook2011, suppC:gao2017}. In addition, they are important
for the understanding of graphene growth
\cite{suppC:kantorovich2017,suppC:luo2020} and are promising
candidates for quantum dot technology \cite{suppC:li2012}. Graphene
nanoislands show much variety in both shape and size, with the favored
shapes governed by their registry and mismatch with the substrate.
Several studies have investigated the structures of small to
medium sized graphene islands (C$_n$ with $n<=24$) on various transition metal surfaces \cite{suppC:lizzit2009, suppC:li2011, suppC:ding2011_1, suppC:ding2011_2, suppC:yang2012, suppC:ding2015, suppC:kantorovich2017}, however in all cases, the exploration of the configurational space was limited to a small number of manually constructed candidates, for each island size. In order to support and extend this body of research, we have, in the present work, used GOFEE to carry out unbiased searches for the lowest energy C$_{18}$ islands on the Ir(111) surface. We choose Ir(111) for its ability to support a particularly high quality graphene layer.
For this problem, GOFEE was applied with two layers of iridium atoms
all fixed at bulk positions and the positions of all carbon atoms were
optimized, starting from randomly initialized positions. To avoid
unintended infinite carbon structures extending through the periodic boundary conditions, carbon atoms were constrained to stay at least $\SI{1}{\angstrom}$ from the edge of the periodic, computational cell, in the plane of the surface. 
During the search, energy and force evaluations were carried out using the GPAW code in LCAO mode with a dzp basis set. Generalized gradient approximation (GGA) with the dispersion corrected optPBE-vdW functional was used to describe the exchange–correlation interaction. Subsequently, the best structures were transferred to a slab with four layers, and relaxed with only the bottom two layers fixed.

\begin{figure*}
	\centering
	\includegraphics[width=0.9\linewidth]{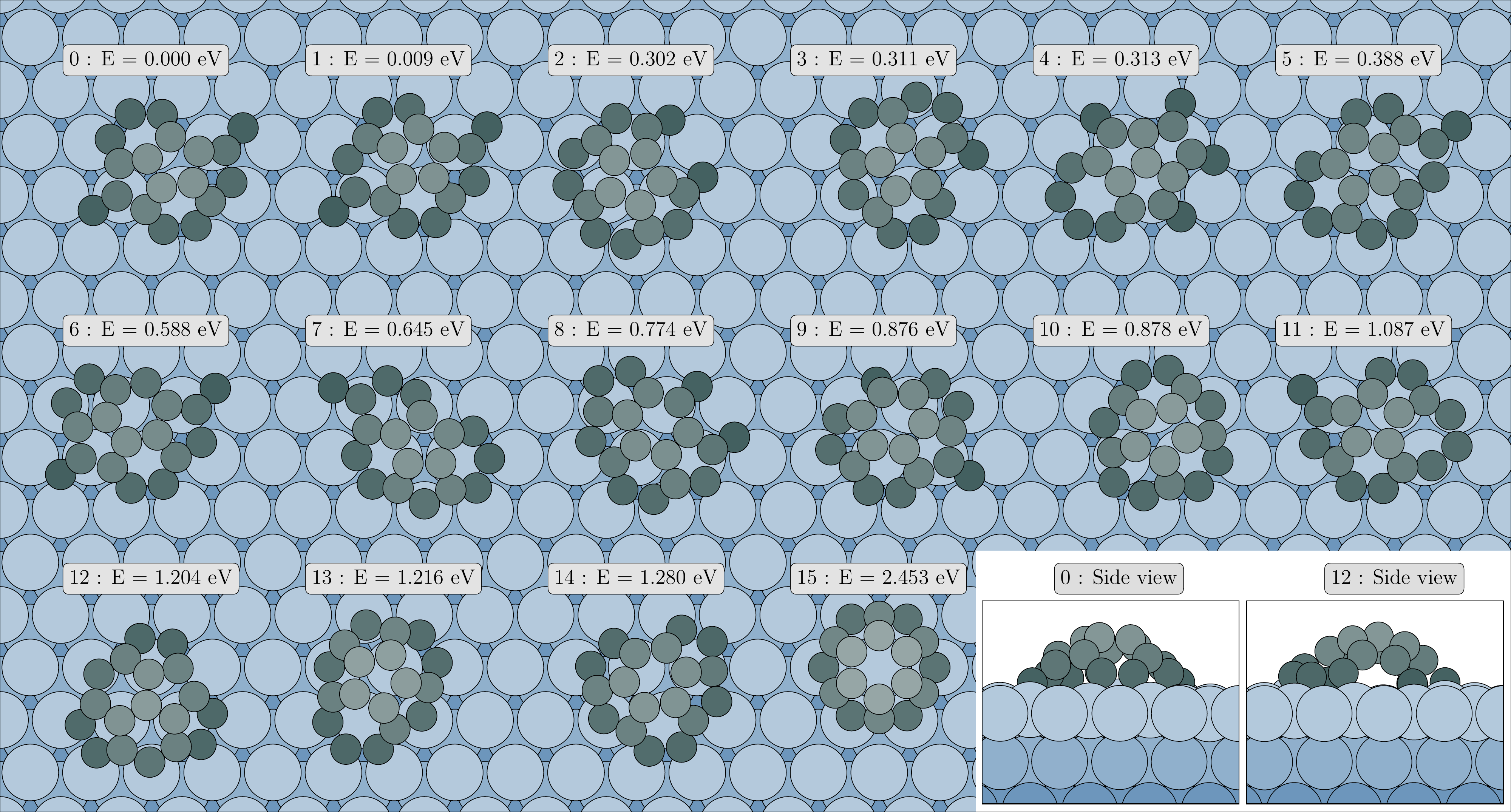}
	\caption{Top views of the lowest energy structures of C$_{18}$
		supported on Ir(111). Carbon and iridium atoms are shown in
		grey and blue colors, respectively. The brightness of the
		colors follow the coordinate perpendicular to the
		surface. Side views are further shown for structures 0 and 4.}
	\label{fig:c18}
\end{figure*}

The resulting lowest energy C$_{18}$/Ir(111) structures are shown in \cref{fig:c18}. All structures exhibit a dome like shape, with the edge atoms bonding to the surface, and the center atoms detached. The predominant building blocks are hexagonal and pentagonal rings, as is also the case for medium to large carbon clusters in the gas phase. However, compared to isolated clusters, the geometry of supported clusters are also governed by their ability to bond to the support. In the structures found, this is represented by a few cases (4,11,13) of heptagonal rings, but especially by the many structures (0-9,11) featuring single atoms branching off closed rings, to bond in hollow sites of the iridium surface. Such branching atoms are present in all structures found within $\approx\SI{0.85}{\electronvolt}$ of the proposed global minimum.
The most predominant edge motives are however still a combination of
closed pentagonal and hexagonal rings. In situations where closed
rings alone presumably do not offer a favorable bonding to the substrate along the entire edge, the results suggest that the favored alternative is the single branching atoms.
The ability of the branching atoms to flexibly bond to the surface is illustrated by a comparison of structures 7 and 9, which feature carbon atoms branching at very different angles to reach the nearest hollow site. 

The complexity and diversity of the structures found is striking and highlights the need for an automated and unbiased search strategy as the present one. We speculate that any heuristic or otherwise human inspired search strategy \cite{suppC:kantorovich2017} would generally fail in identifying all such structural candidates for similar types of problems.


\subsection*{Conclusion}
In conclusion we have here further documented the GOFEE search method \cite{GOFEE}.
This includes how the introduction of an additional, shorter length scale in the kernel can improve short scale resolution and uncertainty estimates of kernel based surrogate models, how the degree of exploration can conveniently be changed and the dramatic improvements resulting from increasing the number of new candidate structures generated in each search iteration. In addition we have presented a new scheme for maintaining a diverse population containing low energy structures as well as a scheme to improve the local optimization of structures during the search.
The search method has been applied to various systems, including surface reconstruction, isolated clusters and supported clusters.

The underlying python implementation is available at \href{http://gofee.au.dk}{gofee.au.dk}.

We acknowledge support from VILLUM FONDEN (Investigator grant, Project No.\ 16562).
This work has been supported by the Danish National Research
Foundation through the Center of Excellence “InterCat” (Grant
agreement no.: DNRF150)

\bibliographystyle{apsrev4-1}
\bibliography{references}

\begin{thebibliography}{84}%
\makeatletter
\providecommand \@ifxundefined [1]{%
 \@ifx{#1\undefined}
}%
\providecommand \@ifnum [1]{%
 \ifnum #1\expandafter \@firstoftwo
 \else \expandafter \@secondoftwo
 \fi
}%
\providecommand \@ifx [1]{%
 \ifx #1\expandafter \@firstoftwo
 \else \expandafter \@secondoftwo
 \fi
}%
\providecommand \natexlab [1]{#1}%
\providecommand \enquote  [1]{``#1''}%
\providecommand \bibnamefont  [1]{#1}%
\providecommand \bibfnamefont [1]{#1}%
\providecommand \citenamefont [1]{#1}%
\providecommand \href@noop [0]{\@secondoftwo}%
\providecommand \href [0]{\begingroup \@sanitize@url \@href}%
\providecommand \@href[1]{\@@startlink{#1}\@@href}%
\providecommand \@@href[1]{\endgroup#1\@@endlink}%
\providecommand \@sanitize@url [0]{\catcode `\\12\catcode `\$12\catcode
  `\&12\catcode `\#12\catcode `\^12\catcode `\_12\catcode `\%12\relax}%
\providecommand \@@startlink[1]{}%
\providecommand \@@endlink[0]{}%
\providecommand \url  [0]{\begingroup\@sanitize@url \@url }%
\providecommand \@url [1]{\endgroup\@href {#1}{\urlprefix }}%
\providecommand \urlprefix  [0]{URL }%
\providecommand \Eprint [0]{\href }%
\providecommand \doibase [0]{http://dx.doi.org/}%
\providecommand \selectlanguage [0]{\@gobble}%
\providecommand \bibinfo  [0]{\@secondoftwo}%
\providecommand \bibfield  [0]{\@secondoftwo}%
\providecommand \translation [1]{[#1]}%
\providecommand \BibitemOpen [0]{}%
\providecommand \bibitemStop [0]{}%
\providecommand \bibitemNoStop [0]{.\EOS\space}%
\providecommand \EOS [0]{\spacefactor3000\relax}%
\providecommand \BibitemShut  [1]{\csname bibitem#1\endcsname}%
\let\auto@bib@innerbib\@empty
\bibitem [{\citenamefont {Pickard}\ and\ \citenamefont
  {Needs}(2011)}]{random_search}%
  \BibitemOpen
  \bibfield  {author} {\bibinfo {author} {\bibfnamefont {C.~J.}\ \bibnamefont
  {Pickard}}\ and\ \bibinfo {author} {\bibfnamefont {R.~J.}\ \bibnamefont
  {Needs}},\ }\href {\doibase 10.1088/0953-8984/23/5/053201} {\bibfield
  {journal} {\bibinfo  {journal} {Journal of Physics: Condensed Matter}\
  }\textbf {\bibinfo {volume} {23}},\ \bibinfo {pages} {053201} (\bibinfo
  {year} {2011})}\BibitemShut {NoStop}%
\bibitem [{\citenamefont {Oganov}\ and\ \citenamefont
  {Glass}(2006)}]{EA:oganov}%
  \BibitemOpen
  \bibfield  {author} {\bibinfo {author} {\bibfnamefont {A.}~\bibnamefont
  {Oganov}}\ and\ \bibinfo {author} {\bibfnamefont {C.}~\bibnamefont {Glass}},\
  }\href {\doibase 10.1063/1.2210932} {\bibfield  {journal} {\bibinfo
  {journal} {The Journal of chemical physics}\ }\textbf {\bibinfo {volume}
  {124}},\ \bibinfo {pages} {244704} (\bibinfo {year} {2006})}\BibitemShut
  {NoStop}%
\bibitem [{\citenamefont {Kolsbjerg}\ \emph {et~al.}(2018)\citenamefont
  {Kolsbjerg}, \citenamefont {Peterson},\ and\ \citenamefont {Hammer}}]{LEA}%
  \BibitemOpen
  \bibfield  {author} {\bibinfo {author} {\bibfnamefont {E.~L.}\ \bibnamefont
  {Kolsbjerg}}, \bibinfo {author} {\bibfnamefont {A.~A.}\ \bibnamefont
  {Peterson}}, \ and\ \bibinfo {author} {\bibfnamefont {B.}~\bibnamefont
  {Hammer}},\ }\href {\doibase 10.1103/PhysRevB.97.195424} {\bibfield
  {journal} {\bibinfo  {journal} {Phys. Rev. B}\ }\textbf {\bibinfo {volume}
  {97}},\ \bibinfo {pages} {195424} (\bibinfo {year} {2018})}\BibitemShut
  {NoStop}%
\bibitem [{\citenamefont {Wales}\ and\ \citenamefont
  {Doye}(1998)}]{basin_hopping}%
  \BibitemOpen
  \bibfield  {author} {\bibinfo {author} {\bibfnamefont {D.}~\bibnamefont
  {Wales}}\ and\ \bibinfo {author} {\bibfnamefont {J.}~\bibnamefont {Doye}},\
  }\href {https://doi.org/10.1021/jp970984n} {\bibfield  {journal} {\bibinfo
  {journal} {The Journal of Physical Chemistry A}\ }\textbf {\bibinfo {volume}
  {101}} (\bibinfo {year} {1998})}\BibitemShut {NoStop}%
\bibitem [{\citenamefont {Goedecker}(2004)}]{minima_hopping}%
  \BibitemOpen
  \bibfield  {author} {\bibinfo {author} {\bibfnamefont {S.}~\bibnamefont
  {Goedecker}},\ }\href {\doibase 10.1063/1.1724816} {\bibfield  {journal}
  {\bibinfo  {journal} {The Journal of chemical physics}\ }\textbf {\bibinfo
  {volume} {120}},\ \bibinfo {pages} {9911} (\bibinfo {year}
  {2004})}\BibitemShut {NoStop}%
\bibitem [{\citenamefont {Wang}\ \emph {et~al.}(2010)\citenamefont {Wang},
  \citenamefont {Lv}, \citenamefont {Zhū},\ and\ \citenamefont
  {Ma}}]{particle_swarm:wang}%
  \BibitemOpen
  \bibfield  {author} {\bibinfo {author} {\bibfnamefont {Y.}~\bibnamefont
  {Wang}}, \bibinfo {author} {\bibfnamefont {J.}~\bibnamefont {Lv}}, \bibinfo
  {author} {\bibfnamefont {L.}~\bibnamefont {Zhū}}, \ and\ \bibinfo {author}
  {\bibfnamefont {Y.}~\bibnamefont {Ma}},\ }\href
  {https://doi.org/10.1103/PhysRevB.82.094116} {\bibfield  {journal} {\bibinfo
  {journal} {Physical Review B}\ }\textbf {\bibinfo {volume} {82}} (\bibinfo
  {year} {2010})}\BibitemShut {NoStop}%
\bibitem [{\citenamefont {Wang}\ \emph {et~al.}(2012)\citenamefont {Wang},
  \citenamefont {Lv}, \citenamefont {Zhū},\ and\ \citenamefont
  {Ma}}]{particle_swarm:CALYPSO}%
  \BibitemOpen
  \bibfield  {author} {\bibinfo {author} {\bibfnamefont {Y.}~\bibnamefont
  {Wang}}, \bibinfo {author} {\bibfnamefont {J.}~\bibnamefont {Lv}}, \bibinfo
  {author} {\bibfnamefont {L.}~\bibnamefont {Zhū}}, \ and\ \bibinfo {author}
  {\bibfnamefont {Y.}~\bibnamefont {Ma}},\ }\href
  {https://doi.org/10.1016/j.cpc.2012.05.008} {\bibfield  {journal} {\bibinfo
  {journal} {Computer Physics Communications}\ }\textbf {\bibinfo {volume}
  {183}} (\bibinfo {year} {2012})}\BibitemShut {NoStop}%
\bibitem [{\citenamefont {Johnston}(2003)}]{EA:johnston}%
  \BibitemOpen
  \bibfield  {author} {\bibinfo {author} {\bibfnamefont {R.}~\bibnamefont
  {Johnston}},\ }\href {https://doi.org/10.1039/b305686d} {\bibfield  {journal}
  {\bibinfo  {journal} {Dalton Transactions - DALTON TRANS}\ }\textbf {\bibinfo
  {volume} {22}} (\bibinfo {year} {2003})}\BibitemShut {NoStop}%
\bibitem [{\citenamefont {Vilhelmsen}\ and\ \citenamefont
  {Hammer}(2014)}]{EA:hammer}%
  \BibitemOpen
  \bibfield  {author} {\bibinfo {author} {\bibfnamefont {L.}~\bibnamefont
  {Vilhelmsen}}\ and\ \bibinfo {author} {\bibfnamefont {B.}~\bibnamefont
  {Hammer}},\ }\href {\doibase 10.1063/1.4886337} {\bibfield  {journal}
  {\bibinfo  {journal} {The Journal of chemical physics}\ }\textbf {\bibinfo
  {volume} {141}},\ \bibinfo {pages} {044711} (\bibinfo {year}
  {2014})}\BibitemShut {NoStop}%
\bibitem [{\citenamefont {Curtis}\ \emph {et~al.}(2018)\citenamefont {Curtis},
  \citenamefont {Li}, \citenamefont {Rose}, \citenamefont
  {V\'azquez-Mayagoitia}, \citenamefont {Bhattacharya}, \citenamefont
  {Ghiringhelli},\ and\ \citenamefont {Marom}}]{EA:GATOR}%
  \BibitemOpen
  \bibfield  {author} {\bibinfo {author} {\bibfnamefont {F.}~\bibnamefont
  {Curtis}}, \bibinfo {author} {\bibfnamefont {X.}~\bibnamefont {Li}}, \bibinfo
  {author} {\bibfnamefont {T.}~\bibnamefont {Rose}}, \bibinfo {author}
  {\bibfnamefont {A.}~\bibnamefont {V\'azquez-Mayagoitia}}, \bibinfo {author}
  {\bibfnamefont {S.}~\bibnamefont {Bhattacharya}}, \bibinfo {author}
  {\bibfnamefont {L.}~\bibnamefont {Ghiringhelli}}, \ and\ \bibinfo {author}
  {\bibfnamefont {N.}~\bibnamefont {Marom}},\ }\href
  {https://doi.org/10.1021/acs.jctc.7b01152} {\bibfield  {journal} {\bibinfo
  {journal} {Journal of Chemical Theory and Computation}\ }\textbf {\bibinfo
  {volume} {14}} (\bibinfo {year} {2018})}\BibitemShut {NoStop}%
\bibitem [{\citenamefont {Lazzeri}\ and\ \citenamefont
  {Selloni}(2001)}]{reconstruction:TiO2}%
  \BibitemOpen
  \bibfield  {author} {\bibinfo {author} {\bibfnamefont {M.}~\bibnamefont
  {Lazzeri}}\ and\ \bibinfo {author} {\bibfnamefont {A.}~\bibnamefont
  {Selloni}},\ }\href {\doibase 10.1103/PhysRevLett.87.266105} {\bibfield
  {journal} {\bibinfo  {journal} {Phys. Rev. Lett.}\ }\textbf {\bibinfo
  {volume} {87}},\ \bibinfo {pages} {266105} (\bibinfo {year}
  {2001})}\BibitemShut {NoStop}%
\bibitem [{\citenamefont {Chuang}\ \emph {et~al.}(2005)\citenamefont {Chuang},
  \citenamefont {Ciobanu}, \citenamefont {Predescu}, \citenamefont {Wang},\
  and\ \citenamefont {Ho}}]{reconstruction:Si114}%
  \BibitemOpen
  \bibfield  {author} {\bibinfo {author} {\bibfnamefont {F.}~\bibnamefont
  {Chuang}}, \bibinfo {author} {\bibfnamefont {C.}~\bibnamefont {Ciobanu}},
  \bibinfo {author} {\bibfnamefont {C.}~\bibnamefont {Predescu}}, \bibinfo
  {author} {\bibfnamefont {C.}~\bibnamefont {Wang}}, \ and\ \bibinfo {author}
  {\bibfnamefont {K.}~\bibnamefont {Ho}},\ }\href
  {https://doi.org/10.1016/j.susc.2005.01.036} {\bibfield  {journal} {\bibinfo
  {journal} {Surface Science}\ }\textbf {\bibinfo {volume} {578}},\ \bibinfo
  {pages} {183} (\bibinfo {year} {2005})}\BibitemShut {NoStop}%
\bibitem [{\citenamefont {Deacon-Smith}\ \emph {et~al.}(2014)\citenamefont
  {Deacon-Smith}, \citenamefont {Scanlon}, \citenamefont {Catlow},
  \citenamefont {Sokol},\ and\ \citenamefont {Woodley}}]{reconstruction:KTaO3}%
  \BibitemOpen
  \bibfield  {author} {\bibinfo {author} {\bibfnamefont {D.~E.~E.}\
  \bibnamefont {Deacon-Smith}}, \bibinfo {author} {\bibfnamefont {D.~O.}\
  \bibnamefont {Scanlon}}, \bibinfo {author} {\bibfnamefont {C.~R.~A.}\
  \bibnamefont {Catlow}}, \bibinfo {author} {\bibfnamefont {A.~A.}\
  \bibnamefont {Sokol}}, \ and\ \bibinfo {author} {\bibfnamefont {S.~M.}\
  \bibnamefont {Woodley}},\ }\href {\doibase
  https://doi.org/10.1002/adma.201401858} {\bibfield  {journal} {\bibinfo
  {journal} {Advanced Materials}\ }\textbf {\bibinfo {volume} {26}},\ \bibinfo
  {pages} {7252} (\bibinfo {year} {2014})}\BibitemShut {NoStop}%
\bibitem [{\citenamefont {Merte}\ \emph {et~al.}(2017)\citenamefont {Merte},
  \citenamefont {J\o{}rgensen}, \citenamefont {Pussi}, \citenamefont
  {Gustafson}, \citenamefont {Shipilin}, \citenamefont {Schaefer},
  \citenamefont {Zhang}, \citenamefont {Rawle}, \citenamefont {Nicklin},
  \citenamefont {Thornton}, \citenamefont {Lindsay}, \citenamefont {Hammer},\
  and\ \citenamefont {Lundgren}}]{SnO2}%
  \BibitemOpen
  \bibfield  {author} {\bibinfo {author} {\bibfnamefont {L.~R.}\ \bibnamefont
  {Merte}}, \bibinfo {author} {\bibfnamefont {M.~S.}\ \bibnamefont
  {J\o{}rgensen}}, \bibinfo {author} {\bibfnamefont {K.}~\bibnamefont {Pussi}},
  \bibinfo {author} {\bibfnamefont {J.}~\bibnamefont {Gustafson}}, \bibinfo
  {author} {\bibfnamefont {M.}~\bibnamefont {Shipilin}}, \bibinfo {author}
  {\bibfnamefont {A.}~\bibnamefont {Schaefer}}, \bibinfo {author}
  {\bibfnamefont {C.}~\bibnamefont {Zhang}}, \bibinfo {author} {\bibfnamefont
  {J.}~\bibnamefont {Rawle}}, \bibinfo {author} {\bibfnamefont
  {C.}~\bibnamefont {Nicklin}}, \bibinfo {author} {\bibfnamefont
  {G.}~\bibnamefont {Thornton}}, \bibinfo {author} {\bibfnamefont
  {R.}~\bibnamefont {Lindsay}}, \bibinfo {author} {\bibfnamefont
  {B.}~\bibnamefont {Hammer}}, \ and\ \bibinfo {author} {\bibfnamefont
  {E.}~\bibnamefont {Lundgren}},\ }\href {\doibase
  10.1103/PhysRevLett.119.096102} {\bibfield  {journal} {\bibinfo  {journal}
  {Phys. Rev. Lett.}\ }\textbf {\bibinfo {volume} {119}},\ \bibinfo {pages}
  {096102} (\bibinfo {year} {2017})}\BibitemShut {NoStop}%
\bibitem [{\citenamefont {M~van~der Zande}\ \emph {et~al.}(2013)\citenamefont
  {M~van~der Zande}, \citenamefont {Y~Huang}, \citenamefont {A~Chenet},
  \citenamefont {C~Berkelbach}, \citenamefont {You}, \citenamefont {Lee},
  \citenamefont {Heinz}, \citenamefont {R~Reichman}, \citenamefont {A~Muller},\
  and\ \citenamefont {Hone}}]{grain_bnd:zande}%
  \BibitemOpen
  \bibfield  {author} {\bibinfo {author} {\bibfnamefont {A.}~\bibnamefont
  {M~van~der Zande}}, \bibinfo {author} {\bibfnamefont {P.}~\bibnamefont
  {Y~Huang}}, \bibinfo {author} {\bibfnamefont {D.}~\bibnamefont {A~Chenet}},
  \bibinfo {author} {\bibfnamefont {T.}~\bibnamefont {C~Berkelbach}}, \bibinfo
  {author} {\bibfnamefont {Y.}~\bibnamefont {You}}, \bibinfo {author}
  {\bibfnamefont {G.-H.}\ \bibnamefont {Lee}}, \bibinfo {author} {\bibfnamefont
  {T.}~\bibnamefont {Heinz}}, \bibinfo {author} {\bibfnamefont
  {D.}~\bibnamefont {R~Reichman}}, \bibinfo {author} {\bibfnamefont
  {D.}~\bibnamefont {A~Muller}}, \ and\ \bibinfo {author} {\bibfnamefont
  {J.}~\bibnamefont {Hone}},\ }\href {\doibase 10.1038/nmat3633} {\bibfield
  {journal} {\bibinfo  {journal} {Nature materials}\ }\textbf {\bibinfo
  {volume} {12}},\ \bibinfo {pages} {554} (\bibinfo {year} {2013})}\BibitemShut
  {NoStop}%
\bibitem [{\citenamefont {Li}\ \emph {et~al.}(2014)\citenamefont {Li},
  \citenamefont {Li}, \citenamefont {Cao}, \citenamefont {Yang}, \citenamefont
  {Shu}, \citenamefont {Zhang}, \citenamefont {Xiang},\ and\ \citenamefont
  {Gong}}]{grain_bnd:xiaowu}%
  \BibitemOpen
  \bibfield  {author} {\bibinfo {author} {\bibfnamefont {Z.-L.}\ \bibnamefont
  {Li}}, \bibinfo {author} {\bibfnamefont {Z.-M.}\ \bibnamefont {Li}}, \bibinfo
  {author} {\bibfnamefont {H.}~\bibnamefont {Cao}}, \bibinfo {author}
  {\bibfnamefont {J.-H.}\ \bibnamefont {Yang}}, \bibinfo {author}
  {\bibfnamefont {Q.}~\bibnamefont {Shu}}, \bibinfo {author} {\bibfnamefont
  {Y.}~\bibnamefont {Zhang}}, \bibinfo {author} {\bibfnamefont
  {H.}~\bibnamefont {Xiang}}, \ and\ \bibinfo {author} {\bibfnamefont
  {X.}~\bibnamefont {Gong}},\ }\href {https://doi.org/10.1039/c3nr06823d}
  {\bibfield  {journal} {\bibinfo  {journal} {Nanoscale}\ }\textbf {\bibinfo
  {volume} {6}} (\bibinfo {year} {2014})}\BibitemShut {NoStop}%
\bibitem [{\citenamefont {Flikkema}\ and\ \citenamefont
  {Bromley}(2004)}]{binary:SiO2}%
  \BibitemOpen
  \bibfield  {author} {\bibinfo {author} {\bibfnamefont {E.}~\bibnamefont
  {Flikkema}}\ and\ \bibinfo {author} {\bibfnamefont {S.}~\bibnamefont
  {Bromley}},\ }\href {\doibase 10.1021/jp049783r} {\bibfield  {journal}
  {\bibinfo  {journal} {J. Phys. Chem. B}\ }\textbf {\bibinfo {volume} {108}},\
  \bibinfo {pages} {9638} (\bibinfo {year} {2004})}\BibitemShut {NoStop}%
\bibitem [{\citenamefont {Ferrando}\ \emph {et~al.}(2008)\citenamefont
  {Ferrando}, \citenamefont {Jellinek},\ and\ \citenamefont
  {Johnston}}]{binary:review}%
  \BibitemOpen
  \bibfield  {author} {\bibinfo {author} {\bibfnamefont {R.}~\bibnamefont
  {Ferrando}}, \bibinfo {author} {\bibfnamefont {J.}~\bibnamefont {Jellinek}},
  \ and\ \bibinfo {author} {\bibfnamefont {R.~L.}\ \bibnamefont {Johnston}},\
  }\href {\doibase 10.1021/cr040090g} {\bibfield  {journal} {\bibinfo
  {journal} {Chem. Rev.}\ }\textbf {\bibinfo {volume} {108}},\ \bibinfo {pages}
  {845} (\bibinfo {year} {2008})}\BibitemShut {NoStop}%
\bibitem [{\citenamefont {Demiroglu}\ \emph {et~al.}(2017)\citenamefont
  {Demiroglu}, \citenamefont {Yao}, \citenamefont {Hussein},\ and\
  \citenamefont {Johnston}}]{cluster_isolated:RuPt}%
  \BibitemOpen
  \bibfield  {author} {\bibinfo {author} {\bibfnamefont {I.}~\bibnamefont
  {Demiroglu}}, \bibinfo {author} {\bibfnamefont {K.}~\bibnamefont {Yao}},
  \bibinfo {author} {\bibfnamefont {H.}~\bibnamefont {Hussein}}, \ and\
  \bibinfo {author} {\bibfnamefont {R.}~\bibnamefont {Johnston}},\ }\href
  {https://doi.org/10.1021/acs.jpcc.6b11329} {\bibfield  {journal} {\bibinfo
  {journal} {The Journal of Physical Chemistry C}\ }\textbf {\bibinfo {volume}
  {121}} (\bibinfo {year} {2017})}\BibitemShut {NoStop}%
\bibitem [{\citenamefont {Aslan}\ \emph {et~al.}(2016)\citenamefont {Aslan},
  \citenamefont {Davis},\ and\ \citenamefont
  {Johnston}}]{cluster:isolated:PdCo}%
  \BibitemOpen
  \bibfield  {author} {\bibinfo {author} {\bibfnamefont {M.}~\bibnamefont
  {Aslan}}, \bibinfo {author} {\bibfnamefont {J.}~\bibnamefont {Davis}}, \ and\
  \bibinfo {author} {\bibfnamefont {R.}~\bibnamefont {Johnston}},\ }\href
  {https://doi.org/10.1039/C6CP00342G} {\bibfield  {journal} {\bibinfo
  {journal} {Phys. Chem. Chem. Phys.}\ }\textbf {\bibinfo {volume} {18}}
  (\bibinfo {year} {2016})}\BibitemShut {NoStop}%
\bibitem [{\citenamefont {Davis}\ \emph {et~al.}(2013)\citenamefont {Davis},
  \citenamefont {Horswell},\ and\ \citenamefont
  {Johnston}}]{cluster_isolated:PdIr}%
  \BibitemOpen
  \bibfield  {author} {\bibinfo {author} {\bibfnamefont {J.}~\bibnamefont
  {Davis}}, \bibinfo {author} {\bibfnamefont {S.}~\bibnamefont {Horswell}}, \
  and\ \bibinfo {author} {\bibfnamefont {R.}~\bibnamefont {Johnston}},\ }\href
  {https://doi.org/10.1021/jp408519z} {\bibfield  {journal} {\bibinfo
  {journal} {The journal of physical chemistry. A}\ }\textbf {\bibinfo {volume}
  {118}} (\bibinfo {year} {2013})}\BibitemShut {NoStop}%
\bibitem [{\citenamefont {Zhai}\ and\ \citenamefont
  {Alexandrova}(2018)}]{clusters:alexandrova1}%
  \BibitemOpen
  \bibfield  {author} {\bibinfo {author} {\bibfnamefont {H.}~\bibnamefont
  {Zhai}}\ and\ \bibinfo {author} {\bibfnamefont {A.}~\bibnamefont
  {Alexandrova}},\ }\href {\doibase 10.1021/acs.jpclett.8b00379} {\bibfield
  {journal} {\bibinfo  {journal} {J. Phys. Chem. Lett.}\ }\textbf {\bibinfo
  {volume} {9}},\ \bibinfo {pages} {1696} (\bibinfo {year} {2018})}\BibitemShut
  {NoStop}%
\bibitem [{\citenamefont {Zandkarimi}\ and\ \citenamefont
  {Alexandrova}(2019)}]{clusters:alexandrova2}%
  \BibitemOpen
  \bibfield  {author} {\bibinfo {author} {\bibfnamefont {B.}~\bibnamefont
  {Zandkarimi}}\ and\ \bibinfo {author} {\bibfnamefont {A.}~\bibnamefont
  {Alexandrova}},\ }\href {\doibase 10.1002/wcms.1420} {\bibfield  {journal}
  {\bibinfo  {journal} {WIREs Comput. Mol. Sci.}\ }\textbf {\bibinfo {volume}
  {0}},\ \bibinfo {pages} {e1420} (\bibinfo {year} {2019})}\BibitemShut
  {NoStop}%
\bibitem [{\citenamefont {Paleico}\ and\ \citenamefont
  {Behler}(2020)}]{clusters:ZnO}%
  \BibitemOpen
  \bibfield  {author} {\bibinfo {author} {\bibfnamefont {M.~L.}\ \bibnamefont
  {Paleico}}\ and\ \bibinfo {author} {\bibfnamefont {J.}~\bibnamefont
  {Behler}},\ }\href {\doibase 10.1063/5.0014876} {\bibfield  {journal}
  {\bibinfo  {journal} {The Journal of Chemical Physics}\ }\textbf {\bibinfo
  {volume} {153}},\ \bibinfo {pages} {054704} (\bibinfo {year}
  {2020})}\BibitemShut {NoStop}%
\bibitem [{\citenamefont {Oganov}\ \emph {et~al.}(2019)\citenamefont {Oganov},
  \citenamefont {Pickard}, \citenamefont {Zhu},\ and\ \citenamefont
  {Needs}}]{solids:review2019}%
  \BibitemOpen
  \bibfield  {author} {\bibinfo {author} {\bibfnamefont {A.}~\bibnamefont
  {Oganov}}, \bibinfo {author} {\bibfnamefont {C.}~\bibnamefont {Pickard}},
  \bibinfo {author} {\bibfnamefont {Q.}~\bibnamefont {Zhu}}, \ and\ \bibinfo
  {author} {\bibfnamefont {R.}~\bibnamefont {Needs}},\ }\href
  {https://doi.org/10.1038/s41578-019-0101-8} {\bibfield  {journal} {\bibinfo
  {journal} {Nature Reviews Materials}\ }\textbf {\bibinfo {volume} {4}}
  (\bibinfo {year} {2019})}\BibitemShut {NoStop}%
\bibitem [{\citenamefont {Bart\'ok}\ \emph {et~al.}(2010)\citenamefont
  {Bart\'ok}, \citenamefont {Payne}, \citenamefont {Kondor},\ and\
  \citenamefont {Cs\'anyi}}]{GAP}%
  \BibitemOpen
  \bibfield  {author} {\bibinfo {author} {\bibfnamefont {A.~P.}\ \bibnamefont
  {Bart\'ok}}, \bibinfo {author} {\bibfnamefont {M.~C.}\ \bibnamefont {Payne}},
  \bibinfo {author} {\bibfnamefont {R.}~\bibnamefont {Kondor}}, \ and\ \bibinfo
  {author} {\bibfnamefont {G.}~\bibnamefont {Cs\'anyi}},\ }\href {\doibase
  10.1103/PhysRevLett.104.136403} {\bibfield  {journal} {\bibinfo  {journal}
  {Phys. Rev. Lett.}\ }\textbf {\bibinfo {volume} {104}},\ \bibinfo {pages}
  {136403} (\bibinfo {year} {2010})}\BibitemShut {NoStop}%
\bibitem [{\citenamefont {Chmiela}\ \emph {et~al.}(2017)\citenamefont
  {Chmiela}, \citenamefont {Tkatchenko}, \citenamefont {Sauceda}, \citenamefont
  {Poltavsky}, \citenamefont {Sch{\"u}tt},\ and\ \citenamefont
  {M{\"u}ller}}]{FF:GDML}%
  \BibitemOpen
  \bibfield  {author} {\bibinfo {author} {\bibfnamefont {S.}~\bibnamefont
  {Chmiela}}, \bibinfo {author} {\bibfnamefont {A.}~\bibnamefont {Tkatchenko}},
  \bibinfo {author} {\bibfnamefont {H.~E.}\ \bibnamefont {Sauceda}}, \bibinfo
  {author} {\bibfnamefont {I.}~\bibnamefont {Poltavsky}}, \bibinfo {author}
  {\bibfnamefont {K.~T.}\ \bibnamefont {Sch{\"u}tt}}, \ and\ \bibinfo {author}
  {\bibfnamefont {K.-R.}\ \bibnamefont {M{\"u}ller}},\ }\href {\doibase
  10.1126/sciadv.1603015} {\bibfield  {journal} {\bibinfo  {journal} {Science
  Advances}\ }\textbf {\bibinfo {volume} {3}},\ \bibinfo {pages} {e1603015}
  (\bibinfo {year} {2017})}\BibitemShut {NoStop}%
\bibitem [{\citenamefont {Behler}\ and\ \citenamefont
  {Parrinello}(2007)}]{parrinello_behler}%
  \BibitemOpen
  \bibfield  {author} {\bibinfo {author} {\bibfnamefont {J.}~\bibnamefont
  {Behler}}\ and\ \bibinfo {author} {\bibfnamefont {M.}~\bibnamefont
  {Parrinello}},\ }\href {\doibase 10.1103/PhysRevLett.98.146401} {\bibfield
  {journal} {\bibinfo  {journal} {Phys. Rev. Lett.}\ }\textbf {\bibinfo
  {volume} {98}},\ \bibinfo {pages} {146401} (\bibinfo {year}
  {2007})}\BibitemShut {NoStop}%
\bibitem [{\citenamefont {Valle}\ and\ \citenamefont
  {Oganov}(2010)}]{oganov_valle}%
  \BibitemOpen
  \bibfield  {author} {\bibinfo {author} {\bibfnamefont {M.}~\bibnamefont
  {Valle}}\ and\ \bibinfo {author} {\bibfnamefont {A.}~\bibnamefont {Oganov}},\
  }\href {\doibase 10.1107/S0108767310026395} {\bibfield  {journal} {\bibinfo
  {journal} {Acta crystallographica. Section A, Foundations of
  crystallography}\ }\textbf {\bibinfo {volume} {66}},\ \bibinfo {pages} {507}
  (\bibinfo {year} {2010})}\BibitemShut {NoStop}%
\bibitem [{\citenamefont {Bartok}\ \emph {et~al.}(2012)\citenamefont {Bartok},
  \citenamefont {Kondor},\ and\ \citenamefont {Csányi}}]{SOAP}%
  \BibitemOpen
  \bibfield  {author} {\bibinfo {author} {\bibfnamefont {A.}~\bibnamefont
  {Bartok}}, \bibinfo {author} {\bibfnamefont {R.}~\bibnamefont {Kondor}}, \
  and\ \bibinfo {author} {\bibfnamefont {G.}~\bibnamefont {Csányi}},\ }\href
  {https://doi.org/10.1103/PhysRevB.87.184115} {\bibfield  {journal} {\bibinfo
  {journal} {Physical Review B}\ }\textbf {\bibinfo {volume} {87}} (\bibinfo
  {year} {2012})}\BibitemShut {NoStop}%
\bibitem [{\citenamefont {Shapeev}(2016)}]{MTP}%
  \BibitemOpen
  \bibfield  {author} {\bibinfo {author} {\bibfnamefont {A.}~\bibnamefont
  {Shapeev}},\ }\href {\doibase 10.1137/15M1054183} {\bibfield  {journal}
  {\bibinfo  {journal} {Multiscale Modeling \& Simulation}\ }\textbf {\bibinfo
  {volume} {14}},\ \bibinfo {pages} {1153} (\bibinfo {year}
  {2016})}\BibitemShut {NoStop}%
\bibitem [{\citenamefont {van~der Oord}\ \emph {et~al.}(2020)\citenamefont
  {van~der Oord}, \citenamefont {Dusson}, \citenamefont {Csányi},\ and\
  \citenamefont {Ortner}}]{FF:ortner}%
  \BibitemOpen
  \bibfield  {author} {\bibinfo {author} {\bibfnamefont {C.}~\bibnamefont
  {van~der Oord}}, \bibinfo {author} {\bibfnamefont {G.}~\bibnamefont
  {Dusson}}, \bibinfo {author} {\bibfnamefont {G.}~\bibnamefont {Csányi}}, \
  and\ \bibinfo {author} {\bibfnamefont {C.}~\bibnamefont {Ortner}},\ }\href
  {https://iopscience.iop.org/article/10.1088/2632-2153/ab527c} {\bibfield
  {journal} {\bibinfo  {journal} {Mach. Learn.: Sci. Technol.}\ }\textbf
  {\bibinfo {volume} {1}},\ \bibinfo {pages} {015004} (\bibinfo {year}
  {2020})}\BibitemShut {NoStop}%
\bibitem [{\citenamefont {Sch{\"u}tt}\ \emph {et~al.}(2018)\citenamefont
  {Sch{\"u}tt}, \citenamefont {Sauceda}, \citenamefont {Kindermans},
  \citenamefont {Tkatchenko},\ and\ \citenamefont {M{\"u}ller}}]{schnet}%
  \BibitemOpen
  \bibfield  {author} {\bibinfo {author} {\bibfnamefont {K.~T.}\ \bibnamefont
  {Sch{\"u}tt}}, \bibinfo {author} {\bibfnamefont {H.~E.}\ \bibnamefont
  {Sauceda}}, \bibinfo {author} {\bibfnamefont {P.-J.}\ \bibnamefont
  {Kindermans}}, \bibinfo {author} {\bibfnamefont {A.}~\bibnamefont
  {Tkatchenko}}, \ and\ \bibinfo {author} {\bibfnamefont {K.-R.}\ \bibnamefont
  {M{\"u}ller}},\ }\href {\doibase 10.1063/1.5019779} {\bibfield  {journal}
  {\bibinfo  {journal} {J. Chem. Phys.}\ }\textbf {\bibinfo {volume} {148}},\
  \bibinfo {pages} {241722} (\bibinfo {year} {2018})}\BibitemShut {NoStop}%
\bibitem [{\citenamefont {Deringer}\ \emph
  {et~al.}(2018{\natexlab{a}})\citenamefont {Deringer}, \citenamefont
  {Pickard},\ and\ \citenamefont {Cs\'anyi}}]{GAPRSS:boron}%
  \BibitemOpen
  \bibfield  {author} {\bibinfo {author} {\bibfnamefont {V.~L.}\ \bibnamefont
  {Deringer}}, \bibinfo {author} {\bibfnamefont {C.~J.}\ \bibnamefont
  {Pickard}}, \ and\ \bibinfo {author} {\bibfnamefont {G.}~\bibnamefont
  {Cs\'anyi}},\ }\href {\doibase 10.1103/PhysRevLett.120.156001} {\bibfield
  {journal} {\bibinfo  {journal} {Phys. Rev. Lett.}\ }\textbf {\bibinfo
  {volume} {120}},\ \bibinfo {pages} {156001} (\bibinfo {year}
  {2018}{\natexlab{a}})}\BibitemShut {NoStop}%
\bibitem [{\citenamefont {Deringer}\ \emph
  {et~al.}(2018{\natexlab{b}})\citenamefont {Deringer}, \citenamefont
  {Proserpio}, \citenamefont {Cs\'anyi},\ and\ \citenamefont
  {Pickard}}]{GAPRSS:crystal}%
  \BibitemOpen
  \bibfield  {author} {\bibinfo {author} {\bibfnamefont {V.~L.}\ \bibnamefont
  {Deringer}}, \bibinfo {author} {\bibfnamefont {D.~M.}\ \bibnamefont
  {Proserpio}}, \bibinfo {author} {\bibfnamefont {G.}~\bibnamefont {Cs\'anyi}},
  \ and\ \bibinfo {author} {\bibfnamefont {C.~J.}\ \bibnamefont {Pickard}},\
  }\href {\doibase 10.1039/C8FD00034D} {\bibfield  {journal} {\bibinfo
  {journal} {Faraday Discuss.}\ }\textbf {\bibinfo {volume} {211}},\ \bibinfo
  {pages} {45} (\bibinfo {year} {2018}{\natexlab{b}})}\BibitemShut {NoStop}%
\bibitem [{\citenamefont {Smith}\ \emph {et~al.}(2018)\citenamefont {Smith},
  \citenamefont {Nebgen}, \citenamefont {Lubbers}, \citenamefont {Isayev},\
  and\ \citenamefont {Roitberg}}]{activeFF:roitberg}%
  \BibitemOpen
  \bibfield  {author} {\bibinfo {author} {\bibfnamefont {J.~S.}\ \bibnamefont
  {Smith}}, \bibinfo {author} {\bibfnamefont {B.}~\bibnamefont {Nebgen}},
  \bibinfo {author} {\bibfnamefont {N.}~\bibnamefont {Lubbers}}, \bibinfo
  {author} {\bibfnamefont {O.}~\bibnamefont {Isayev}}, \ and\ \bibinfo {author}
  {\bibfnamefont {A.~E.}\ \bibnamefont {Roitberg}},\ }\href {\doibase
  10.1063/1.5023802} {\bibfield  {journal} {\bibinfo  {journal} {J. Chem.
  Phys.}\ }\textbf {\bibinfo {volume} {148}},\ \bibinfo {pages} {241733}
  (\bibinfo {year} {2018})}\BibitemShut {NoStop}%
\bibitem [{\citenamefont {Zhang}\ \emph {et~al.}(2019)\citenamefont {Zhang},
  \citenamefont {Lin}, \citenamefont {Wang}, \citenamefont {Car},\ and\
  \citenamefont {E}}]{activeFF:Zhang}%
  \BibitemOpen
  \bibfield  {author} {\bibinfo {author} {\bibfnamefont {L.}~\bibnamefont
  {Zhang}}, \bibinfo {author} {\bibfnamefont {D.-Y.}\ \bibnamefont {Lin}},
  \bibinfo {author} {\bibfnamefont {H.}~\bibnamefont {Wang}}, \bibinfo {author}
  {\bibfnamefont {R.}~\bibnamefont {Car}}, \ and\ \bibinfo {author}
  {\bibfnamefont {W.}~\bibnamefont {E}},\ }\href {\doibase
  10.1103/PhysRevMaterials.3.023804} {\bibfield  {journal} {\bibinfo  {journal}
  {Physical Review Materials}\ }\textbf {\bibinfo {volume} {3}},\ \bibinfo
  {pages} {023804} (\bibinfo {year} {2019})}\BibitemShut {NoStop}%
\bibitem [{\citenamefont {Gubaev}\ \emph {et~al.}(2017)\citenamefont {Gubaev},
  \citenamefont {Podryabinkin},\ and\ \citenamefont
  {Shapeev}}]{active:shapeev2017}%
  \BibitemOpen
  \bibfield  {author} {\bibinfo {author} {\bibfnamefont {K.}~\bibnamefont
  {Gubaev}}, \bibinfo {author} {\bibfnamefont {E.}~\bibnamefont
  {Podryabinkin}}, \ and\ \bibinfo {author} {\bibfnamefont {A.}~\bibnamefont
  {Shapeev}},\ }\href {https://doi.org/10.1063/1.5005095} {\bibfield  {journal}
  {\bibinfo  {journal} {The Journal of Chemical Physics}\ }\textbf {\bibinfo
  {volume} {148}} (\bibinfo {year} {2017})}\BibitemShut {NoStop}%
\bibitem [{\citenamefont {Schran}\ \emph {et~al.}(2020)\citenamefont {Schran},
  \citenamefont {Brezina},\ and\ \citenamefont {Marsalek}}]{active:ondrej2020}%
  \BibitemOpen
  \bibfield  {author} {\bibinfo {author} {\bibfnamefont {C.}~\bibnamefont
  {Schran}}, \bibinfo {author} {\bibfnamefont {K.}~\bibnamefont {Brezina}}, \
  and\ \bibinfo {author} {\bibfnamefont {O.}~\bibnamefont {Marsalek}},\ }\href
  {\doibase 10.1063/5.0016004} {\bibfield  {journal} {\bibinfo  {journal} {The
  Journal of Chemical Physics}\ }\textbf {\bibinfo {volume} {153}},\ \bibinfo
  {pages} {104105} (\bibinfo {year} {2020})}\BibitemShut {NoStop}%
\bibitem [{\citenamefont {Smith}\ \emph {et~al.}(2020)\citenamefont {Smith},
  \citenamefont {Zubatyuk}, \citenamefont {Nebgen}, \citenamefont {Lubbers},
  \citenamefont {Barros}, \citenamefont {Roitberg}, \citenamefont {Isayev},\
  and\ \citenamefont {Tretiak}}]{active:smith2020}%
  \BibitemOpen
  \bibfield  {author} {\bibinfo {author} {\bibfnamefont {J.~S.}\ \bibnamefont
  {Smith}}, \bibinfo {author} {\bibfnamefont {R.}~\bibnamefont {Zubatyuk}},
  \bibinfo {author} {\bibfnamefont {B.~T.}\ \bibnamefont {Nebgen}}, \bibinfo
  {author} {\bibfnamefont {N.}~\bibnamefont {Lubbers}}, \bibinfo {author}
  {\bibfnamefont {K.}~\bibnamefont {Barros}}, \bibinfo {author} {\bibfnamefont
  {A.}~\bibnamefont {Roitberg}}, \bibinfo {author} {\bibfnamefont
  {O.}~\bibnamefont {Isayev}}, \ and\ \bibinfo {author} {\bibfnamefont
  {S.}~\bibnamefont {Tretiak}},\ }\href {\doibase
  https://doi.org/10.1038/s41597-020-0473-z} {\bibfield  {journal} {\bibinfo
  {journal} {Scientific Data}\ }\textbf {\bibinfo {volume} {7}},\ \bibinfo
  {pages} {134} (\bibinfo {year} {2020})}\BibitemShut {NoStop}%
\bibitem [{\citenamefont {Doan}\ \emph {et~al.}(2020)\citenamefont {Doan},
  \citenamefont {Agarwal}, \citenamefont {Qian}, \citenamefont {Counihan},
  \citenamefont {Rodríguez-López}, \citenamefont {Moore},\ and\ \citenamefont
  {Assary}}]{active:assary2020}%
  \BibitemOpen
  \bibfield  {author} {\bibinfo {author} {\bibfnamefont {H.~A.}\ \bibnamefont
  {Doan}}, \bibinfo {author} {\bibfnamefont {G.}~\bibnamefont {Agarwal}},
  \bibinfo {author} {\bibfnamefont {H.}~\bibnamefont {Qian}}, \bibinfo {author}
  {\bibfnamefont {M.~J.}\ \bibnamefont {Counihan}}, \bibinfo {author}
  {\bibfnamefont {J.}~\bibnamefont {Rodríguez-López}}, \bibinfo {author}
  {\bibfnamefont {J.~S.}\ \bibnamefont {Moore}}, \ and\ \bibinfo {author}
  {\bibfnamefont {R.~S.}\ \bibnamefont {Assary}},\ }\href
  {https://doi.org/10.1021/acs.chemmater.0c00768} {\bibfield  {journal}
  {\bibinfo  {journal} {Chemistry of Materials}\ }\textbf {\bibinfo {volume}
  {32}},\ \bibinfo {pages} {6338} (\bibinfo {year} {2020})}\BibitemShut
  {NoStop}%
\bibitem [{\citenamefont {Garijo~del R\'{\i}o}\ \emph
  {et~al.}(2019)\citenamefont {Garijo~del R\'{\i}o}, \citenamefont
  {Mortensen},\ and\ \citenamefont {Jacobsen}}]{local:karsten}%
  \BibitemOpen
  \bibfield  {author} {\bibinfo {author} {\bibfnamefont {E.}~\bibnamefont
  {Garijo~del R\'{\i}o}}, \bibinfo {author} {\bibfnamefont {J.~J.}\
  \bibnamefont {Mortensen}}, \ and\ \bibinfo {author} {\bibfnamefont {K.~W.}\
  \bibnamefont {Jacobsen}},\ }\href {\doibase 10.1103/PhysRevB.100.104103}
  {\bibfield  {journal} {\bibinfo  {journal} {Phys. Rev. B}\ }\textbf {\bibinfo
  {volume} {100}},\ \bibinfo {pages} {104103} (\bibinfo {year}
  {2019})}\BibitemShut {NoStop}%
\bibitem [{\citenamefont {Denzel}\ and\ \citenamefont
  {K{\"a}stner}(2018)}]{local:kastner}%
  \BibitemOpen
  \bibfield  {author} {\bibinfo {author} {\bibfnamefont {A.}~\bibnamefont
  {Denzel}}\ and\ \bibinfo {author} {\bibfnamefont {J.}~\bibnamefont
  {K{\"a}stner}},\ }\href {\doibase 10.1063/1.5017103} {\bibfield  {journal}
  {\bibinfo  {journal} {J. Chem. Phys.}\ }\textbf {\bibinfo {volume} {148}},\
  \bibinfo {pages} {094114} (\bibinfo {year} {2018})}\BibitemShut {NoStop}%
\bibitem [{\citenamefont {Peterson}(2016)}]{neb:andrew2016}%
  \BibitemOpen
  \bibfield  {author} {\bibinfo {author} {\bibfnamefont {A.~A.}\ \bibnamefont
  {Peterson}},\ }\href {\doibase 10.1063/1.4960708} {\bibfield  {journal}
  {\bibinfo  {journal} {J. Chem. Phys.}\ }\textbf {\bibinfo {volume} {145}},\
  \bibinfo {pages} {074106} (\bibinfo {year} {2016})}\BibitemShut {NoStop}%
\bibitem [{\citenamefont {Koistinen}\ \emph {et~al.}(2017)\citenamefont
  {Koistinen}, \citenamefont {Dagbjartsd\'{o}ttir}, \citenamefont
  {\'{A}sgeirsson}, \citenamefont {Vehtari},\ and\ \citenamefont
  {J\'{o}nsson}}]{neb:hannes2017}%
  \BibitemOpen
  \bibfield  {author} {\bibinfo {author} {\bibfnamefont {O.-P.}\ \bibnamefont
  {Koistinen}}, \bibinfo {author} {\bibfnamefont {F.~B.}\ \bibnamefont
  {Dagbjartsd\'{o}ttir}}, \bibinfo {author} {\bibfnamefont {V.}~\bibnamefont
  {\'{A}sgeirsson}}, \bibinfo {author} {\bibfnamefont {A.}~\bibnamefont
  {Vehtari}}, \ and\ \bibinfo {author} {\bibfnamefont {H.}~\bibnamefont
  {J\'{o}nsson}},\ }\href {\doibase 10.1063/1.4986787} {\bibfield  {journal}
  {\bibinfo  {journal} {J. Chem. Phys.}\ }\textbf {\bibinfo {volume} {147}},\
  \bibinfo {pages} {152720} (\bibinfo {year} {2017})}\BibitemShut {NoStop}%
\bibitem [{\citenamefont {Garrido~Torres}\ \emph {et~al.}(2019)\citenamefont
  {Garrido~Torres}, \citenamefont {Jennings}, \citenamefont {Hansen},
  \citenamefont {Boes},\ and\ \citenamefont {Bligaard}}]{neb:bligaard2019}%
  \BibitemOpen
  \bibfield  {author} {\bibinfo {author} {\bibfnamefont {J.~A.}\ \bibnamefont
  {Garrido~Torres}}, \bibinfo {author} {\bibfnamefont {P.~C.}\ \bibnamefont
  {Jennings}}, \bibinfo {author} {\bibfnamefont {M.~H.}\ \bibnamefont
  {Hansen}}, \bibinfo {author} {\bibfnamefont {J.~R.}\ \bibnamefont {Boes}}, \
  and\ \bibinfo {author} {\bibfnamefont {T.}~\bibnamefont {Bligaard}},\ }\href
  {\doibase 10.1103/PhysRevLett.122.156001} {\bibfield  {journal} {\bibinfo
  {journal} {Phys. Rev. Lett.}\ }\textbf {\bibinfo {volume} {122}},\ \bibinfo
  {pages} {156001} (\bibinfo {year} {2019})}\BibitemShut {NoStop}%
\bibitem [{\citenamefont {Li}\ \emph {et~al.}(2015)\citenamefont {Li},
  \citenamefont {Kermode},\ and\ \citenamefont {De~Vita}}]{activeMD:Vita}%
  \BibitemOpen
  \bibfield  {author} {\bibinfo {author} {\bibfnamefont {Z.}~\bibnamefont
  {Li}}, \bibinfo {author} {\bibfnamefont {J.~R.}\ \bibnamefont {Kermode}}, \
  and\ \bibinfo {author} {\bibfnamefont {A.}~\bibnamefont {De~Vita}},\ }\href
  {\doibase 10.1103/PhysRevLett.114.096405} {\bibfield  {journal} {\bibinfo
  {journal} {Phys. Rev. Lett.}\ }\textbf {\bibinfo {volume} {114}},\ \bibinfo
  {pages} {096405} (\bibinfo {year} {2015})}\BibitemShut {NoStop}%
\bibitem [{\citenamefont {A.~Peterson}\ \emph {et~al.}(2017)\citenamefont
  {A.~Peterson}, \citenamefont {Christensen},\ and\ \citenamefont
  {Khorshidi}}]{activeMD:andrew}%
  \BibitemOpen
  \bibfield  {author} {\bibinfo {author} {\bibfnamefont {A.}~\bibnamefont
  {A.~Peterson}}, \bibinfo {author} {\bibfnamefont {R.}~\bibnamefont
  {Christensen}}, \ and\ \bibinfo {author} {\bibfnamefont {A.}~\bibnamefont
  {Khorshidi}},\ }\href {\doibase 10.1039/C7CP00375G} {\bibfield  {journal}
  {\bibinfo  {journal} {Phys. Chem. Chem. Phys.}\ }\textbf {\bibinfo {volume}
  {19}},\ \bibinfo {pages} {10978} (\bibinfo {year} {2017})}\BibitemShut
  {NoStop}%
\bibitem [{\citenamefont {Miwa}\ and\ \citenamefont
  {Ohno}(2017)}]{activeMD:Ohno}%
  \BibitemOpen
  \bibfield  {author} {\bibinfo {author} {\bibfnamefont {K.}~\bibnamefont
  {Miwa}}\ and\ \bibinfo {author} {\bibfnamefont {H.}~\bibnamefont {Ohno}},\
  }\href {\doibase 10.1103/PhysRevMaterials.1.053801} {\bibfield  {journal}
  {\bibinfo  {journal} {Phys. Rev. Materials}\ }\textbf {\bibinfo {volume}
  {1}},\ \bibinfo {pages} {053801} (\bibinfo {year} {2017})}\BibitemShut
  {NoStop}%
\bibitem [{\citenamefont {Podryabinkin}\ and\ \citenamefont
  {Shapeev}(2017)}]{activeMD:evgeny}%
  \BibitemOpen
  \bibfield  {author} {\bibinfo {author} {\bibfnamefont {E.~V.}\ \bibnamefont
  {Podryabinkin}}\ and\ \bibinfo {author} {\bibfnamefont {A.~V.}\ \bibnamefont
  {Shapeev}},\ }\href {\doibase
  https://doi.org/10.1016/j.commatsci.2017.08.031} {\bibfield  {journal}
  {\bibinfo  {journal} {Comput. Mater. Sci.}\ }\textbf {\bibinfo {volume}
  {140}},\ \bibinfo {pages} {171 } (\bibinfo {year} {2017})}\BibitemShut
  {NoStop}%
\bibitem [{\citenamefont {Novikov}\ \emph {et~al.}(2018)\citenamefont
  {Novikov}, \citenamefont {Suleimanov},\ and\ \citenamefont
  {Shapeev}}]{activeMD:Shapeev}%
  \BibitemOpen
  \bibfield  {author} {\bibinfo {author} {\bibfnamefont {I.}~\bibnamefont
  {Novikov}}, \bibinfo {author} {\bibfnamefont {Y.}~\bibnamefont {Suleimanov}},
  \ and\ \bibinfo {author} {\bibfnamefont {A.}~\bibnamefont {Shapeev}},\ }\href
  {https://doi.org/10.1039/C8CP06037A} {\bibfield  {journal} {\bibinfo
  {journal} {Physical Chemistry Chemical Physics}\ }\textbf {\bibinfo {volume}
  {20}} (\bibinfo {year} {2018})}\BibitemShut {NoStop}%
\bibitem [{\citenamefont {Jinnouchi}\ \emph {et~al.}(2019)\citenamefont
  {Jinnouchi}, \citenamefont {Karsai},\ and\ \citenamefont
  {Kresse}}]{activeMD:kresse}%
  \BibitemOpen
  \bibfield  {author} {\bibinfo {author} {\bibfnamefont {R.}~\bibnamefont
  {Jinnouchi}}, \bibinfo {author} {\bibfnamefont {F.}~\bibnamefont {Karsai}}, \
  and\ \bibinfo {author} {\bibfnamefont {G.}~\bibnamefont {Kresse}},\ }\href
  {\doibase 10.1103/PhysRevB.100.014105} {\bibfield  {journal} {\bibinfo
  {journal} {Phys. Rev. B}\ }\textbf {\bibinfo {volume} {100}},\ \bibinfo
  {pages} {014105} (\bibinfo {year} {2019})}\BibitemShut {NoStop}%
\bibitem [{\citenamefont {Ulissi}\ \emph {et~al.}(2017)\citenamefont {Ulissi},
  \citenamefont {Medford}, \citenamefont {Bligaard},\ and\ \citenamefont
  {Nørskov}}]{activeReact:Noerskov}%
  \BibitemOpen
  \bibfield  {author} {\bibinfo {author} {\bibfnamefont {Z.}~\bibnamefont
  {Ulissi}}, \bibinfo {author} {\bibfnamefont {A.}~\bibnamefont {Medford}},
  \bibinfo {author} {\bibfnamefont {T.}~\bibnamefont {Bligaard}}, \ and\
  \bibinfo {author} {\bibfnamefont {J.}~\bibnamefont {Nørskov}},\ }\href
  {\doibase 10.1038/ncomms14621} {\bibfield  {journal} {\bibinfo  {journal}
  {Nature Communications}\ }\textbf {\bibinfo {volume} {8}},\ \bibinfo {pages}
  {14621} (\bibinfo {year} {2017})}\BibitemShut {NoStop}%
\bibitem [{\citenamefont {Stocker}\ \emph {et~al.}(2020)\citenamefont
  {Stocker}, \citenamefont {Csányi}, \citenamefont {Reuter},\ and\
  \citenamefont {Margraf}}]{activeReact:Stocker}%
  \BibitemOpen
  \bibfield  {author} {\bibinfo {author} {\bibfnamefont {S.}~\bibnamefont
  {Stocker}}, \bibinfo {author} {\bibfnamefont {G.}~\bibnamefont {Csányi}},
  \bibinfo {author} {\bibfnamefont {K.}~\bibnamefont {Reuter}}, \ and\ \bibinfo
  {author} {\bibfnamefont {J.}~\bibnamefont {Margraf}},\ }\href
  {https://doi.org/10.1038/s41467-020-19267-x} {\bibfield  {journal} {\bibinfo
  {journal} {Nature Communications}\ }\textbf {\bibinfo {volume} {11}}
  (\bibinfo {year} {2020})}\BibitemShut {NoStop}%
\bibitem [{\citenamefont {Zhai}\ \emph {et~al.}(2015)\citenamefont {Zhai},
  \citenamefont {Ha},\ and\ \citenamefont {Alexandrova}}]{SS:alexandrova}%
  \BibitemOpen
  \bibfield  {author} {\bibinfo {author} {\bibfnamefont {H.}~\bibnamefont
  {Zhai}}, \bibinfo {author} {\bibfnamefont {M.-A.}\ \bibnamefont {Ha}}, \ and\
  \bibinfo {author} {\bibfnamefont {A.~N.}\ \bibnamefont {Alexandrova}},\
  }\href {\doibase 10.1021/acs.jctc.5b00065} {\bibfield  {journal} {\bibinfo
  {journal} {J. Chem. Theory Comput.}\ }\textbf {\bibinfo {volume} {11}},\
  \bibinfo {pages} {2385} (\bibinfo {year} {2015})}\BibitemShut {NoStop}%
\bibitem [{\citenamefont {Todorovi{\'c}}\ \emph {et~al.}(2017)\citenamefont
  {Todorovi{\'c}}, \citenamefont {Gutmann}, \citenamefont {Corander},\ and\
  \citenamefont {Rinke}}]{active:rinke}%
  \BibitemOpen
  \bibfield  {author} {\bibinfo {author} {\bibfnamefont {M.}~\bibnamefont
  {Todorovi{\'c}}}, \bibinfo {author} {\bibfnamefont {M.}~\bibnamefont
  {Gutmann}}, \bibinfo {author} {\bibfnamefont {J.}~\bibnamefont {Corander}}, \
  and\ \bibinfo {author} {\bibfnamefont {P.}~\bibnamefont {Rinke}},\ }\href
  {\doibase 10.1038/s41524-019-0175-2} {\bibfield  {journal} {\bibinfo
  {journal} {npj Computational Materials}\ }\textbf {\bibinfo {volume} {5}},\
  \bibinfo {pages} {35} (\bibinfo {year} {2017})}\BibitemShut {NoStop}%
\bibitem [{\citenamefont {Yamashita}\ \emph {et~al.}(2018)\citenamefont
  {Yamashita}, \citenamefont {Sato}, \citenamefont {Kino}, \citenamefont
  {Miyake}, \citenamefont {Tsuda},\ and\ \citenamefont
  {Oguchi}}]{active:oguchi}%
  \BibitemOpen
  \bibfield  {author} {\bibinfo {author} {\bibfnamefont {T.}~\bibnamefont
  {Yamashita}}, \bibinfo {author} {\bibfnamefont {N.}~\bibnamefont {Sato}},
  \bibinfo {author} {\bibfnamefont {H.}~\bibnamefont {Kino}}, \bibinfo {author}
  {\bibfnamefont {T.}~\bibnamefont {Miyake}}, \bibinfo {author} {\bibfnamefont
  {K.}~\bibnamefont {Tsuda}}, \ and\ \bibinfo {author} {\bibfnamefont
  {T.}~\bibnamefont {Oguchi}},\ }\href {\doibase
  10.1103/PhysRevMaterials.2.013803} {\bibfield  {journal} {\bibinfo  {journal}
  {Phys. Rev. Materials}\ }\textbf {\bibinfo {volume} {2}},\ \bibinfo {pages}
  {013803} (\bibinfo {year} {2018})}\BibitemShut {NoStop}%
\bibitem [{\citenamefont {Tong}\ \emph {et~al.}(2018)\citenamefont {Tong},
  \citenamefont {Xue}, \citenamefont {Lv}, \citenamefont {Wang},\ and\
  \citenamefont {ma}}]{activeSS:calypso}%
  \BibitemOpen
  \bibfield  {author} {\bibinfo {author} {\bibfnamefont {Q.}~\bibnamefont
  {Tong}}, \bibinfo {author} {\bibfnamefont {L.}~\bibnamefont {Xue}}, \bibinfo
  {author} {\bibfnamefont {J.}~\bibnamefont {Lv}}, \bibinfo {author}
  {\bibfnamefont {Y.}~\bibnamefont {Wang}}, \ and\ \bibinfo {author}
  {\bibfnamefont {Y.}~\bibnamefont {ma}},\ }\href
  {https://doi.org/10.1039/C8FD00055G} {\bibfield  {journal} {\bibinfo
  {journal} {Faraday Discussions}\ }\textbf {\bibinfo {volume} {211}} (\bibinfo
  {year} {2018})}\BibitemShut {NoStop}%
\bibitem [{\citenamefont {Gubaev}\ \emph {et~al.}(2019)\citenamefont {Gubaev},
  \citenamefont {Podryabinkin}, \citenamefont {Hart},\ and\ \citenamefont
  {Shapeev}}]{activeSS:shapeev}%
  \BibitemOpen
  \bibfield  {author} {\bibinfo {author} {\bibfnamefont {K.}~\bibnamefont
  {Gubaev}}, \bibinfo {author} {\bibfnamefont {E.}~\bibnamefont
  {Podryabinkin}}, \bibinfo {author} {\bibfnamefont {G.}~\bibnamefont {Hart}},
  \ and\ \bibinfo {author} {\bibfnamefont {A.}~\bibnamefont {Shapeev}},\ }\href
  {\doibase 10.1016/j.commatsci.2018.09.031} {\bibfield  {journal} {\bibinfo
  {journal} {Comput. Mater. Sci.}\ }\textbf {\bibinfo {volume} {156}},\
  \bibinfo {pages} {148} (\bibinfo {year} {2019})}\BibitemShut {NoStop}%
\bibitem [{\citenamefont {Van~den Bossche}(2019)}]{SS_dftb:maxime}%
  \BibitemOpen
  \bibfield  {author} {\bibinfo {author} {\bibfnamefont {M.}~\bibnamefont
  {Van~den Bossche}},\ }\href {\doibase 10.1021/acs.jpca.9b00927} {\bibfield
  {journal} {\bibinfo  {journal} {J. Phys. Chem. A}\ }\textbf {\bibinfo
  {volume} {123}},\ \bibinfo {pages} {3038} (\bibinfo {year}
  {2019})}\BibitemShut {NoStop}%
\bibitem [{\citenamefont {Louren\c{c}o}\ \emph {et~al.}(2020)\citenamefont
  {Louren\c{c}o}, \citenamefont {Anast\'{a}cio}, \citenamefont {Da~Rosa},
  \citenamefont {Frauenheim},\ and\ \citenamefont {Silva}}]{activeSS:daSilva}%
  \BibitemOpen
  \bibfield  {author} {\bibinfo {author} {\bibfnamefont {M.}~\bibnamefont
  {Louren\c{c}o}}, \bibinfo {author} {\bibfnamefont {A.}~\bibnamefont
  {Anast\'{a}cio}}, \bibinfo {author} {\bibfnamefont {A.~L.}\ \bibnamefont
  {Da~Rosa}}, \bibinfo {author} {\bibfnamefont {T.}~\bibnamefont {Frauenheim}},
  \ and\ \bibinfo {author} {\bibfnamefont {M.}~\bibnamefont {Silva}},\ }\href
  {https://doi.org/10.1007/s00894-020-04438-w} {\bibfield  {journal} {\bibinfo
  {journal} {Journal of Molecular Modeling}\ }\textbf {\bibinfo {volume} {26}}
  (\bibinfo {year} {2020})}\BibitemShut {NoStop}%
\bibitem [{\citenamefont {Bisbo}\ and\ \citenamefont {Hammer}(2020)}]{GOFEE}%
  \BibitemOpen
  \bibfield  {author} {\bibinfo {author} {\bibfnamefont {M.~K.}\ \bibnamefont
  {Bisbo}}\ and\ \bibinfo {author} {\bibfnamefont {B.}~\bibnamefont {Hammer}},\
  }\href {\doibase 10.1103/PhysRevLett.124.086102} {\bibfield  {journal}
  {\bibinfo  {journal} {Phys. Rev. Lett.}\ }\textbf {\bibinfo {volume} {124}},\
  \bibinfo {pages} {086102} (\bibinfo {year} {2020})}\BibitemShut {NoStop}%
\bibitem [{\citenamefont {Dolgonos}\ \emph {et~al.}(2010)\citenamefont
  {Dolgonos}, \citenamefont {Aradi}, \citenamefont {Moreira},\ and\
  \citenamefont {Frauenheim}}]{DFTB:TiO2}%
  \BibitemOpen
  \bibfield  {author} {\bibinfo {author} {\bibfnamefont {G.}~\bibnamefont
  {Dolgonos}}, \bibinfo {author} {\bibfnamefont {B.}~\bibnamefont {Aradi}},
  \bibinfo {author} {\bibfnamefont {N.~H.}\ \bibnamefont {Moreira}}, \ and\
  \bibinfo {author} {\bibfnamefont {T.}~\bibnamefont {Frauenheim}},\ }\href
  {\doibase 10.1021/ct900422c} {\bibfield  {journal} {\bibinfo  {journal}
  {Journal of Chemical Theory and Computation}\ }\textbf {\bibinfo {volume}
  {6}},\ \bibinfo {pages} {266} (\bibinfo {year} {2010})}\BibitemShut {NoStop}%
\bibitem [{\citenamefont {Rasmussen}\ and\ \citenamefont
  {Williams}(2005)}]{GP:Rasmussen}%
  \BibitemOpen
  \bibfield  {author} {\bibinfo {author} {\bibfnamefont {C.~E.}\ \bibnamefont
  {Rasmussen}}\ and\ \bibinfo {author} {\bibfnamefont {C.~K.~I.}\ \bibnamefont
  {Williams}},\ }\href@noop {} {\emph {\bibinfo {title} {Gaussian Processes for
  Machine Learning}}}\ (\bibinfo  {publisher} {The MIT Press},\ \bibinfo {year}
  {2005})\BibitemShut {NoStop}%
\bibitem [{\citenamefont {Mortensen}\ \emph {et~al.}(2020)\citenamefont
  {Mortensen}, \citenamefont {Meldgaard}, \citenamefont {Bisbo}, \citenamefont
  {Christiansen},\ and\ \citenamefont {Hammer}}]{Mortensen_2020}%
  \BibitemOpen
  \bibfield  {author} {\bibinfo {author} {\bibfnamefont {H.~L.}\ \bibnamefont
  {Mortensen}}, \bibinfo {author} {\bibfnamefont {S.~A.}\ \bibnamefont
  {Meldgaard}}, \bibinfo {author} {\bibfnamefont {M.~K.}\ \bibnamefont
  {Bisbo}}, \bibinfo {author} {\bibfnamefont {M.-P.~V.}\ \bibnamefont
  {Christiansen}}, \ and\ \bibinfo {author} {\bibfnamefont {B.}~\bibnamefont
  {Hammer}},\ }\href {\doibase 10.1103/PhysRevB.102.075427} {\bibfield
  {journal} {\bibinfo  {journal} {Phys. Rev. B}\ }\textbf {\bibinfo {volume}
  {102}},\ \bibinfo {pages} {075427} (\bibinfo {year} {2020})}\BibitemShut
  {NoStop}%
\bibitem [{\citenamefont {Jørgensen}\ \emph {et~al.}(2017)\citenamefont
  {Jørgensen}, \citenamefont {Groves},\ and\ \citenamefont
  {Hammer}}]{jorgensen_2017}%
  \BibitemOpen
  \bibfield  {author} {\bibinfo {author} {\bibfnamefont {M.~S.}\ \bibnamefont
  {Jørgensen}}, \bibinfo {author} {\bibfnamefont {M.~N.}\ \bibnamefont
  {Groves}}, \ and\ \bibinfo {author} {\bibfnamefont {B.}~\bibnamefont
  {Hammer}},\ }\href {\doibase 10.1021/acs.jctc.6b01119} {\bibfield  {journal}
  {\bibinfo  {journal} {Journal of Chemical Theory and Computation}\ }\textbf
  {\bibinfo {volume} {13}},\ \bibinfo {pages} {1486} (\bibinfo {year}
  {2017})}\BibitemShut {NoStop}%
\bibitem [{\citenamefont {{Wang}}\ \emph {et~al.}(2017)\citenamefont {{Wang}},
  \citenamefont {{van Stein}}, \citenamefont {{Emmerich}},\ and\ \citenamefont
  {{Back}}}]{acquisition}%
  \BibitemOpen
  \bibfield  {author} {\bibinfo {author} {\bibfnamefont {H.}~\bibnamefont
  {{Wang}}}, \bibinfo {author} {\bibfnamefont {B.}~\bibnamefont {{van Stein}}},
  \bibinfo {author} {\bibfnamefont {M.}~\bibnamefont {{Emmerich}}}, \ and\
  \bibinfo {author} {\bibfnamefont {T.}~\bibnamefont {{Back}}},\ }in\ \href
  {\doibase 10.1109/SMC.2017.8122656} {\emph {\bibinfo {booktitle} {2017 IEEE
  International Conference on Systems, Man, and Cybernetics (SMC)}}}\ (\bibinfo
  {year} {2017})\ pp.\ \bibinfo {pages} {507--512}\BibitemShut {NoStop}%
\bibitem [{\citenamefont {H{\"a}se}\ \emph {et~al.}(2018)\citenamefont
  {H{\"a}se}, \citenamefont {Roch}, \citenamefont {Kreisbeck},\ and\
  \citenamefont {Aspuru-Guzik}}]{phoenics}%
  \BibitemOpen
  \bibfield  {author} {\bibinfo {author} {\bibfnamefont {F.}~\bibnamefont
  {H{\"a}se}}, \bibinfo {author} {\bibfnamefont {L.~M.}\ \bibnamefont {Roch}},
  \bibinfo {author} {\bibfnamefont {C.}~\bibnamefont {Kreisbeck}}, \ and\
  \bibinfo {author} {\bibfnamefont {A.}~\bibnamefont {Aspuru-Guzik}},\ }\href
  {\doibase 10.1021/acscentsci.8b00307} {\bibfield  {journal} {\bibinfo
  {journal} {ACS Central Science}\ }\textbf {\bibinfo {volume} {4}},\ \bibinfo
  {pages} {1134} (\bibinfo {year} {2018})}\BibitemShut {NoStop}%
\bibitem [{\citenamefont {Jørgensen}\ \emph {et~al.}(2018)\citenamefont
  {Jørgensen}, \citenamefont {Larsen}, \citenamefont {Jacobsen},\ and\
  \citenamefont {Hammer}}]{explore:mathias2018}%
  \BibitemOpen
  \bibfield  {author} {\bibinfo {author} {\bibfnamefont {M.~S.}\ \bibnamefont
  {Jørgensen}}, \bibinfo {author} {\bibfnamefont {U.~F.}\ \bibnamefont
  {Larsen}}, \bibinfo {author} {\bibfnamefont {K.~W.}\ \bibnamefont
  {Jacobsen}}, \ and\ \bibinfo {author} {\bibfnamefont {B.}~\bibnamefont
  {Hammer}},\ }\href {\doibase 10.1021/acs.jpca.8b00160} {\bibfield  {journal}
  {\bibinfo  {journal} {The Journal of Physical Chemistry A}\ }\textbf
  {\bibinfo {volume} {122}},\ \bibinfo {pages} {1504} (\bibinfo {year}
  {2018})}\BibitemShut {NoStop}%
\bibitem [{\citenamefont {Caro}(2019)}]{fastSOAP}%
  \BibitemOpen
  \bibfield  {author} {\bibinfo {author} {\bibfnamefont {M.~A.}\ \bibnamefont
  {Caro}},\ }\href {\doibase 10.1103/PhysRevB.100.024112} {\bibfield  {journal}
  {\bibinfo  {journal} {Phys. Rev. B}\ }\textbf {\bibinfo {volume} {100}},\
  \bibinfo {pages} {024112} (\bibinfo {year} {2019})}\BibitemShut {NoStop}%
\bibitem [{\citenamefont {Kocer}\ \emph {et~al.}(2020)\citenamefont {Kocer},
  \citenamefont {Mason},\ and\ \citenamefont {Erturk}}]{descriptor:bessel}%
  \BibitemOpen
  \bibfield  {author} {\bibinfo {author} {\bibfnamefont {E.}~\bibnamefont
  {Kocer}}, \bibinfo {author} {\bibfnamefont {J.}~\bibnamefont {Mason}}, \ and\
  \bibinfo {author} {\bibfnamefont {H.}~\bibnamefont {Erturk}},\ }\href
  {\doibase 10.1063/1.5111045} {\bibfield  {journal} {\bibinfo  {journal} {AIP
  Advances}\ }\textbf {\bibinfo {volume} {10}},\ \bibinfo {pages} {015021}
  (\bibinfo {year} {2020})}\BibitemShut {NoStop}%
\bibitem [{\citenamefont {An}\ \emph {et~al.}(2008)\citenamefont {An},
  \citenamefont {Shao}, \citenamefont {Bulusu},\ and\ \citenamefont
  {Zeng}}]{C24}%
  \BibitemOpen
  \bibfield  {author} {\bibinfo {author} {\bibfnamefont {W.}~\bibnamefont
  {An}}, \bibinfo {author} {\bibfnamefont {N.}~\bibnamefont {Shao}}, \bibinfo
  {author} {\bibfnamefont {S.}~\bibnamefont {Bulusu}}, \ and\ \bibinfo {author}
  {\bibfnamefont {X.~C.}\ \bibnamefont {Zeng}},\ }\href {\doibase
  10.1063/1.2831917} {\bibfield  {journal} {\bibinfo  {journal} {J. Chem.
  Phys.}\ }\textbf {\bibinfo {volume} {128}},\ \bibinfo {pages} {084301}
  (\bibinfo {year} {2008})}\BibitemShut {NoStop}%
\bibitem [{\citenamefont {Meunier}\ \emph {et~al.}(2016)\citenamefont
  {Meunier}, \citenamefont {Souza~Filho}, \citenamefont {Barros},\ and\
  \citenamefont {Dresselhaus}}]{suppC:dresselhaus2016}%
  \BibitemOpen
  \bibfield  {author} {\bibinfo {author} {\bibfnamefont {V.}~\bibnamefont
  {Meunier}}, \bibinfo {author} {\bibfnamefont {A.~G.}\ \bibnamefont
  {Souza~Filho}}, \bibinfo {author} {\bibfnamefont {E.~B.}\ \bibnamefont
  {Barros}}, \ and\ \bibinfo {author} {\bibfnamefont {M.~S.}\ \bibnamefont
  {Dresselhaus}},\ }\href {\doibase 10.1103/RevModPhys.88.025005} {\bibfield
  {journal} {\bibinfo  {journal} {Rev. Mod. Phys.}\ }\textbf {\bibinfo {volume}
  {88}},\ \bibinfo {pages} {025005} (\bibinfo {year} {2016})}\BibitemShut
  {NoStop}%
\bibitem [{\citenamefont {Barnard}\ and\ \citenamefont
  {Snook}(2011)}]{suppC:snook2011}%
  \BibitemOpen
  \bibfield  {author} {\bibinfo {author} {\bibfnamefont {A.}~\bibnamefont
  {Barnard}}\ and\ \bibinfo {author} {\bibfnamefont {I.}~\bibnamefont
  {Snook}},\ }\href {\doibase 10.1088/0965-0393/19/5/054001} {\bibfield
  {journal} {\bibinfo  {journal} {Modelling and Simulation in Materials Science
  and Engineering}\ }\textbf {\bibinfo {volume} {19}},\ \bibinfo {pages}
  {054001} (\bibinfo {year} {2011})}\BibitemShut {NoStop}%
\bibitem [{\citenamefont {Chen}\ \emph {et~al.}(2017)\citenamefont {Chen},
  \citenamefont {Que}, \citenamefont {Tao}, \citenamefont {Zhang},
  \citenamefont {Lin}, \citenamefont {Xiao}, \citenamefont {Wang},
  \citenamefont {Du}, \citenamefont {Pantelides},\ and\ \citenamefont
  {Gao}}]{suppC:gao2017}%
  \BibitemOpen
  \bibfield  {author} {\bibinfo {author} {\bibfnamefont {H.}~\bibnamefont
  {Chen}}, \bibinfo {author} {\bibfnamefont {Y.}~\bibnamefont {Que}}, \bibinfo
  {author} {\bibfnamefont {L.}~\bibnamefont {Tao}}, \bibinfo {author}
  {\bibfnamefont {Y.-Y.}\ \bibnamefont {Zhang}}, \bibinfo {author}
  {\bibfnamefont {X.}~\bibnamefont {Lin}}, \bibinfo {author} {\bibfnamefont
  {W.}~\bibnamefont {Xiao}}, \bibinfo {author} {\bibfnamefont {D.}~\bibnamefont
  {Wang}}, \bibinfo {author} {\bibfnamefont {S.}~\bibnamefont {Du}}, \bibinfo
  {author} {\bibfnamefont {S.}~\bibnamefont {Pantelides}}, \ and\ \bibinfo
  {author} {\bibfnamefont {H.-J.}\ \bibnamefont {Gao}},\ }\href {\doibase
  10.1007/s12274-017-1940-5} {\bibfield  {journal} {\bibinfo  {journal} {Nano
  Research}\ }\textbf {\bibinfo {volume} {11}},\ \bibinfo {pages} {1} (\bibinfo
  {year} {2017})}\BibitemShut {NoStop}%
\bibitem [{\citenamefont {Tetlow}\ \emph {et~al.}(2017)\citenamefont {Tetlow},
  \citenamefont {Ford},\ and\ \citenamefont
  {Kantorovich}}]{suppC:kantorovich2017}%
  \BibitemOpen
  \bibfield  {author} {\bibinfo {author} {\bibfnamefont {H.}~\bibnamefont
  {Tetlow}}, \bibinfo {author} {\bibfnamefont {I.}~\bibnamefont {Ford}}, \ and\
  \bibinfo {author} {\bibfnamefont {L.}~\bibnamefont {Kantorovich}},\ }\href
  {\doibase 10.1063/1.4974335} {\bibfield  {journal} {\bibinfo  {journal} {The
  Journal of Chemical Physics}\ }\textbf {\bibinfo {volume} {146}},\ \bibinfo
  {pages} {044702} (\bibinfo {year} {2017})}\BibitemShut {NoStop}%
\bibitem [{\citenamefont {Liu}\ \emph {et~al.}(2020)\citenamefont {Liu},
  \citenamefont {Xu}, \citenamefont {Zhang}, \citenamefont {Dong},
  \citenamefont {Wang},\ and\ \citenamefont {Luo}}]{suppC:luo2020}%
  \BibitemOpen
  \bibfield  {author} {\bibinfo {author} {\bibfnamefont {Y.}~\bibnamefont
  {Liu}}, \bibinfo {author} {\bibfnamefont {L.}~\bibnamefont {Xu}}, \bibinfo
  {author} {\bibfnamefont {L.}~\bibnamefont {Zhang}}, \bibinfo {author}
  {\bibfnamefont {Z.}~\bibnamefont {Dong}}, \bibinfo {author} {\bibfnamefont
  {S.}~\bibnamefont {Wang}}, \ and\ \bibinfo {author} {\bibfnamefont
  {L.}~\bibnamefont {Luo}},\ }\href {\doibase 10.1021/acsami.0c15990}
  {\bibfield  {journal} {\bibinfo  {journal} {ACS Applied Materials \&
  Interfaces}\ }\textbf {\bibinfo {volume} {12}},\ \bibinfo {pages} {52201}
  (\bibinfo {year} {2020})}\BibitemShut {NoStop}%
\bibitem [{\citenamefont {Shen}\ \emph {et~al.}(2012)\citenamefont {Shen},
  \citenamefont {Zhu}, \citenamefont {Yang},\ and\ \citenamefont
  {Li}}]{suppC:li2012}%
  \BibitemOpen
  \bibfield  {author} {\bibinfo {author} {\bibfnamefont {J.}~\bibnamefont
  {Shen}}, \bibinfo {author} {\bibfnamefont {Y.}~\bibnamefont {Zhu}}, \bibinfo
  {author} {\bibfnamefont {X.}~\bibnamefont {Yang}}, \ and\ \bibinfo {author}
  {\bibfnamefont {C.}~\bibnamefont {Li}},\ }\href {\doibase 10.1039/c2cc00110a}
  {\bibfield  {journal} {\bibinfo  {journal} {Chemical communications
  (Cambridge, England)}\ }\textbf {\bibinfo {volume} {48}},\ \bibinfo {pages}
  {3686} (\bibinfo {year} {2012})}\BibitemShut {NoStop}%
\bibitem [{\citenamefont {Lacovig}\ \emph {et~al.}(2009)\citenamefont
  {Lacovig}, \citenamefont {Pozzo}, \citenamefont {Alf\`e}, \citenamefont
  {Vilmercati}, \citenamefont {Baraldi},\ and\ \citenamefont
  {Lizzit}}]{suppC:lizzit2009}%
  \BibitemOpen
  \bibfield  {author} {\bibinfo {author} {\bibfnamefont {P.}~\bibnamefont
  {Lacovig}}, \bibinfo {author} {\bibfnamefont {M.}~\bibnamefont {Pozzo}},
  \bibinfo {author} {\bibfnamefont {D.}~\bibnamefont {Alf\`e}}, \bibinfo
  {author} {\bibfnamefont {P.}~\bibnamefont {Vilmercati}}, \bibinfo {author}
  {\bibfnamefont {A.}~\bibnamefont {Baraldi}}, \ and\ \bibinfo {author}
  {\bibfnamefont {S.}~\bibnamefont {Lizzit}},\ }\href {\doibase
  10.1103/PhysRevLett.103.166101} {\bibfield  {journal} {\bibinfo  {journal}
  {Phys. Rev. Lett.}\ }\textbf {\bibinfo {volume} {103}},\ \bibinfo {pages}
  {166101} (\bibinfo {year} {2009})}\BibitemShut {NoStop}%
\bibitem [{\citenamefont {Wang}\ \emph {et~al.}(2011)\citenamefont {Wang},
  \citenamefont {Ma}, \citenamefont {Caffio}, \citenamefont {Schaub},\ and\
  \citenamefont {Li}}]{suppC:li2011}%
  \BibitemOpen
  \bibfield  {author} {\bibinfo {author} {\bibfnamefont {B.}~\bibnamefont
  {Wang}}, \bibinfo {author} {\bibfnamefont {X.}~\bibnamefont {Ma}}, \bibinfo
  {author} {\bibfnamefont {M.}~\bibnamefont {Caffio}}, \bibinfo {author}
  {\bibfnamefont {R.}~\bibnamefont {Schaub}}, \ and\ \bibinfo {author}
  {\bibfnamefont {W.-X.}\ \bibnamefont {Li}},\ }\href {\doibase
  10.1021/nl103053t} {\bibfield  {journal} {\bibinfo  {journal} {Nano letters}\
  }\textbf {\bibinfo {volume} {11}},\ \bibinfo {pages} {424} (\bibinfo {year}
  {2011})}\BibitemShut {NoStop}%
\bibitem [{\citenamefont {Gao}\ \emph {et~al.}(2011)\citenamefont {Gao},
  \citenamefont {Yuan}, \citenamefont {Hu}, \citenamefont {Zhao},\ and\
  \citenamefont {Ding}}]{suppC:ding2011_1}%
  \BibitemOpen
  \bibfield  {author} {\bibinfo {author} {\bibfnamefont {J.}~\bibnamefont
  {Gao}}, \bibinfo {author} {\bibfnamefont {Q.}~\bibnamefont {Yuan}}, \bibinfo
  {author} {\bibfnamefont {H.}~\bibnamefont {Hu}}, \bibinfo {author}
  {\bibfnamefont {J.}~\bibnamefont {Zhao}}, \ and\ \bibinfo {author}
  {\bibfnamefont {F.}~\bibnamefont {Ding}},\ }\href {\doibase
  10.1021/jp2051454} {\bibfield  {journal} {\bibinfo  {journal} {The Journal of
  Physical Chemistry C}\ }\textbf {\bibinfo {volume} {115}},\ \bibinfo {pages}
  {17695} (\bibinfo {year} {2011})}\BibitemShut {NoStop}%
\bibitem [{\citenamefont {Yuan}\ \emph {et~al.}(2011)\citenamefont {Yuan},
  \citenamefont {Gao}, \citenamefont {Shu}, \citenamefont {Zhao}, \citenamefont
  {Chen},\ and\ \citenamefont {Ding}}]{suppC:ding2011_2}%
  \BibitemOpen
  \bibfield  {author} {\bibinfo {author} {\bibfnamefont {Q.}~\bibnamefont
  {Yuan}}, \bibinfo {author} {\bibfnamefont {J.}~\bibnamefont {Gao}}, \bibinfo
  {author} {\bibfnamefont {H.}~\bibnamefont {Shu}}, \bibinfo {author}
  {\bibfnamefont {J.}~\bibnamefont {Zhao}}, \bibinfo {author} {\bibfnamefont
  {X.}~\bibnamefont {Chen}}, \ and\ \bibinfo {author} {\bibfnamefont
  {F.}~\bibnamefont {Ding}},\ }\href {\doibase 10.1021/ja2050875} {\bibfield
  {journal} {\bibinfo  {journal} {Journal of the American Chemical Society}\
  }\textbf {\bibinfo {volume} {134}},\ \bibinfo {pages} {2970} (\bibinfo {year}
  {2011})}\BibitemShut {NoStop}%
\bibitem [{\citenamefont {Wu}\ \emph {et~al.}(2012)\citenamefont {Wu},
  \citenamefont {Jiang}, \citenamefont {Zhang}, \citenamefont {Li},
  \citenamefont {Hou},\ and\ \citenamefont {Yang}}]{suppC:yang2012}%
  \BibitemOpen
  \bibfield  {author} {\bibinfo {author} {\bibfnamefont {P.}~\bibnamefont
  {Wu}}, \bibinfo {author} {\bibfnamefont {H.}~\bibnamefont {Jiang}}, \bibinfo
  {author} {\bibfnamefont {W.}~\bibnamefont {Zhang}}, \bibinfo {author}
  {\bibfnamefont {Z.}~\bibnamefont {Li}}, \bibinfo {author} {\bibfnamefont
  {Z.}~\bibnamefont {Hou}}, \ and\ \bibinfo {author} {\bibfnamefont
  {J.}~\bibnamefont {Yang}},\ }\href {\doibase 10.1021/ja301791x} {\bibfield
  {journal} {\bibinfo  {journal} {Journal of the American Chemical Society}\
  }\textbf {\bibinfo {volume} {134}},\ \bibinfo {pages} {6045} (\bibinfo {year}
  {2012})}\BibitemShut {NoStop}%
\bibitem [{\citenamefont {Gao}\ and\ \citenamefont
  {Ding}(2015)}]{suppC:ding2015}%
  \BibitemOpen
  \bibfield  {author} {\bibinfo {author} {\bibfnamefont {J.}~\bibnamefont
  {Gao}}\ and\ \bibinfo {author} {\bibfnamefont {F.}~\bibnamefont {Ding}},\
  }\href {https://doi.org/10.1007/s10876-014-0834-x} {\bibfield  {journal}
  {\bibinfo  {journal} {Journal of Cluster Science}\ }\textbf {\bibinfo
  {volume} {26}} (\bibinfo {year} {2015})}\BibitemShut {NoStop}%
\end{thebibliography}%

\end{document}